\documentclass[twocolumn]{aastex631}

\newcommand{\Nminusone}{N_{\mathrm{+,1.0}}}
\DeclareUnicodeCharacter{2212}{-}
\usepackage{textcomp}
\usepackage{gensymb}
\usepackage{amsmath}
\usepackage{url}
\usepackage{float}

\begin{document}

\title{Standardized Luminosity of the Tip of the Red Giant Branch utilizing Multiple Fields in NGC 4258 and the CATs Algorithm}

\correspondingauthor{Siyang Li}
\email{sli185@jh.edu}

\author[0000-0002-8623-1082]{Siyang Li}
    \affil{Department of Physics and Astronomy, Johns Hopkins University, Baltimore, MD 21218, USA}
    
\author[0000-0002-6124-1196]{Adam G. Riess}
    \affil{Department of Physics and Astronomy, Johns Hopkins University, Baltimore, MD 21218, USA}
    \affil{Space Telescope Science Institute, Baltimore, MD, 21218, USA}    

\author[0000-0002-4934-5849]{Dan Scolnic}
    \affil{Department of Physics, Duke University, Durham, NC 27708, USA}

\author[0000-0002-5259-2314]{Gagandeep S. Anand}
    \affil{Space Telescope Science Institute, Baltimore, MD, 21218, USA}

\author[00000-0003-3829-967X]{Jiaxi Wu}
    \affil{Kuang Yaming Honors School, Nanjing University, Nanjing, Jiangsu 210023, China}
    \affil{Department of Physics, Duke University, Durham, NC 27708, USA}

\author{Stefano Casertano}
    \affil{Space Telescope Science Institute, Baltimore, MD, 21218, USA}
    
\author[0000-0001-9420-6525]{Wenlong Yuan}
\affiliation{Department of Physics and Astronomy, Johns Hopkins University, Baltimore, MD 21218, USA}

\author[0000-0002-1691-8217]{Rachael Beaton}
    \affil{Space Telescope Science Institute, Baltimore, MD, 21218, USA}
    
\author[0000-0001-8089-4419]{Richard I. Anderson}
\affiliation{Institute of Physics, \'Ecole Polytechnique F\'ed\'erale de Lausanne (EPFL), Observatoire de Sauverny, 1290 Versoix, Switzerland}

\begin{abstract}
The Tip of the Red Giant Branch provides a luminous standard candle for calibrating distance ladders that reach Type Ia supernova (SN Ia) hosts.  However, recent work reveals that tip measurements vary at the $\sim$ 0.1 mag level for different stellar populations and locations within a host, which may lead to inconsistencies along the distance ladder.  We pursue a calibration of the tip using 11 \textit{Hubble Space Telescope} fields around the maser host, NGC 4258, that is consistent with SN Ia hosts by standardizing tip measurements via their contrast ratios. We find $F814W$-band tips that exhibit a full 0.3 mag range and 0.1 mag dispersion.  We do not find any correlation between \ion{H}{1} column density and the apparent tip to 0.04 $\pm$ 0.03~mag/cm$^{-2}$. We search for a tip-contrast relation (TCR) and measure the TCR within the fields of NGC 4258 of $-0.015\pm0.008$ mag/$R$, where $R$ is the contrast ratio. This value is consistent with the TCR originally discovered in the GHOSTS sample \citep{Wu_2022arXiv221106354W} of $-0.023\pm0.005$~mag/R.  Combining these measurements, we find a global TCR of $-0.021\pm0.004$~mag/R and a calibration of $M_I^{TRGB} = -4.025 \pm 0.035 - (R-4)\times0.021$~mag. We also use stellar models to simulate single age and metallicity stellar populations with [Fe/H] from $-2.0$ to $-0.7$ and ages from 3 Gyr to 12 Gyr and reconstruct the global TCR found here to a factor of $\sim$ 2.  This work is combined in a companion analysis with tip measurements of nearby SN Ia hosts to measure $H_0$.  
\end{abstract}

\keywords{}

\section{Introduction}
\label{sec:intro}

The Tip of the Red Giant Branch (TRGB) is a near-standard candle that forms as a consequence of the onset of helium burning in the degenerate cores of low mass red giant stars \citep{Serenelli_2017A&A...606A..33S}. Because of its bright luminosity ($M_I \sim -4$~mag), the tip can be used to measure distances to galaxies beyond the Local Group \cite[for instance,][]{Lee_1993ApJ...417..553L, Radburn-Smith_2011ApJS..195...18R, Lee_2012ApJ...760L..14L, Tully_2013AJ....146...86T, Jang_2017ApJ...835...28J, Jang_2021ApJ...906..125J, Anand_EDD_2021AJ....162...80A} and aid in the construction of extragalactic distance ladders to measure the Hubble constant, $H_0$  \citep{Ferrarese_2000ApJ...529..745F, Tammann_2008ApJ...679...52T, Mould_3_2009ApJ...697..996M, Hislop_2011ApJ...733...75H, Lee_2012ApJ...760L..14L, Lee_2013ApJ...773...13L, Tammann_2013A&A...549A.136T, Jang_2015ApJ...807..133J, Freedman_2019ApJ...882...34F, Freedman_2020ApJ...891...57F, Kim_2020ApJ...905..104K, Blakeslee_2021ApJ...911...65B, Freedman_2001ApJ...553...47F, Anand_EDDvsCCHP_2021arXiv210800007A, Anderson_2023arXiv230304790A}.  It has been long known that the tip luminosity varies with color  \citep{Lee_1993ApJ...417..553L, Rizzi_2007ApJ...661..815R, Madore_2009ApJ...690..389M, Jang_2017ApJ...835...28J, McQuinn_2019ApJ...880...63M}, and its color-rectified luminosity or luminosity measured over a modest color range has been treated as a standard candle. 

However, observations have revealed that measurements of the luminosity of the tip varies at the $\sim$0.1~mag level in the I-band even over a small, $\sim$ 1 mag $V-I$ color range \cite[for a compilation and discussion of recent measured luminosities with NGC 4258, Large Magellanic Cloud (LMC), and the Milky Way, see][]{Blakeslee_2021ApJ...911...65B, Freedman_Tensions_2021ApJ...919...16F, Li_2022arXiv220211110L}.   Strong evidence of this variation is anchored by LMC studies.  \cite{Hoyt_2023NatAs.tmp...58H} observed field-to-field variations of several tenths of a magnitude (after correction for extinction) accompanied by differences at the $<0.03$ mag level in the width of the Sobel responses of the luminosity function.  Similarly, \cite{Anderson_2023arXiv230304790A} observed a 0.1 mag level difference in the tip of two sub-populations of red giants (those with small-scale variability in the A or B sequence) with 5 $\sigma$ confidence.  Unfortunately, these metrics (width or variability) require very high signal to noise ratio data that are not readily available for more distant hosts.  They are also not measurable in the maser host, NGC 4258, which serves as perhaps the most important calibrator of the tip due to the ability to observe it with the same facilities and under similar conditions as SN Ia hosts.

Another measure that correlates with tip brightness, but which is feasible to measure in distant hosts, is the tip contrast ratio (ratio of stars below and above the measured tip).  Its relation with the tip has been observed at high (5 $\sigma$) confidence by \cite{Wu_2022arXiv221106354W} by comparing multiple halo pointings for several hosts in the GHOSTs sample which can be attributed in part due to the measurement method itself.  This property, largely driven by the presence of younger, asymptotic giant branch (AGB) stars more luminous than the tip, is expected to be related to characteristics of the stellar populations such as age and metallicity \citep{Wu_2022arXiv221106354W} based on the study of stellar isochrones.  In analyses, it is essential that tip measurements are \textit{standardized} to account for variations that arise from these stellar properties. 

\subsection{TRGB Measurements} 

Empirically, the tip is measured by locating the discontinuity in the number of stars along the giant branch luminosity function, typically computed from inside a diagonal band placed on a color magnitude diagram. This discontinuity can be located using edge detection methods such as a Sobel filter \citep[][and references therein]{Lee_1993ApJ...417..553L,  Sakai_1996ApJ...461..713S, Hatt_2017ApJ...845..146H} or with a multi-parameter model fit to the luminosity function using least-square minimization \citep{Wu_2014AJ....148....7W} or maximum likelihood estimation \citep[][and references therein]{Mendez_2002AJ....124..213M, Makarov_2006AJ....132.2729M,  Mcquinn_2016ApJ...826...21M, Kim_2020ApJ...905..104K, Li_2022arXiv220211110L}. For the edge detection approach, a filter, most often of the Sobel type, is evaluated across the luminosity function to measure its weighted first derivative and identify the TRGB discontinuity \citep[peak in the derivative function;][]{Lee_1993ApJ...417..553L}. However, this approach can result in several local maxima in the first derivative that are comparable in height or which are outside an `expected' range of tip values based on prior inferences. These effects can lead to a somewhat subjective differentiation between the true tip and spurious features.   Previous measurements of the tip in NGC 4258 \citep[such as][]{Madore_2009ApJ...690..389M} involved supervision to decide which tip to accept and subjective definition of what area constitutes ``the halo'' \citep{Jang_2021ApJ...906..125J}.  To address these issues, \cite{Wu_2022arXiv221106354W} developed an unsupervised, Sobel filter-based tip measurement algorithm as part of the Comparative Analysis of tips (CATs) program and optimized its parameters to minimize the dispersion among multiple fields in Galaxy Halos, Outer disks, Star clusters, Thick disks, and Substructure (GHOSTS) survey \citep{Radburn-Smith_2011ApJS..195...18R} galaxies. The CATs algorithm from  \cite{Wu_2022arXiv221106354W} differs from past tip measurement algorithms by using quantifiable, objective procedures that remove some of the subjective steps used in other work such as choice of color band slopes and widths.  It uses an entirely unsupervised Sobel filter-based approach and accepts multiple tip measurements per population; this obviates the need to select a specific peak in the Sobel response, and accounts for variations in stellar properties that can produce tips of slightly different luminosities. The detections undergo a series of quality cuts that are applied in an unsupervised and consistent manner. 

\cite{Wu_2022arXiv221106354W} found that the measured tip magnitude becomes brighter with increasing contrast ratio, where the contrast ratio is defined as the quotient between the number of stars 0.5~mag fainter and brighter than the measured tip and may be related to properties of different stellar populations, luminosity function characteristics, and measurement parameter choices that cause variations in the measured tip. We note that the exact origin of this relationship is inconsequential to the validity of its application as a calibrating tool in general (though the slope may change with smoothing and noise, see Sections \ref{sec:ArtpopTCR} and \ref{sec:Discussion}). The tip-contrast ratio relationship can be used to standardize the tip so that distances to galaxies are calibrated in a more similar manner. When measuring the distance to a galaxy, the measured apparent magnitude tip should be tied to an absolute magnitude tip or zero-point that is calculated using a population of similar stellar properties, luminosity function characteristics, and measurement parameter choices, otherwise the zero-point may not accurately reflect the true absolute magnitude of the tip of a population with those stellar properties and characteristics measured with a different setup. It will be important to investigate whether this relationship exists in other galaxies, especially those used to anchor the extragalactic distance ladder to measure $H_0$. NGC 4258 acts as an anchor galaxy for the TRGB-based distance ladder due to the availability of a geometric, maser-based distance \citep{Pesce_2020ApJ...891L...1P, Reid_2019ApJ...886L..27R} and is an ideal testing ground for this hypothesis as multiple fields have been observed in the past (see Section \ref{sec:Data_Selection}). 

In this study, we use the CATs algorithm to calibrate the tip-contrast ratio relationship in NGC 4258. We describe our data selection procedure in Section \ref{sec:Data_Selection} and first provide an estimate of the average tip measured across all fields in NGC 4258, before any correction, in Section \ref{sec:Multi-field_Average}. We then calibrate the tip-contrast ratio relationship in Section \ref{sec:TCR}. We recreate a tip-contrast ratio relationship with simulated populations in \ref{sec:ArtpopTCR} and discuss our results in Section \ref{sec:Discussion}. 

\section{Data Selection}
\label{sec:Data_Selection}

We retrieve all publicly available images of NGC 4258 in the approximate region of its halo (25th mag isophote) that were taken in the \textit{Hubble Space Telescope} ($HST$) ACS/WFC $F814W$, $F555W$ and $F606W$ filters from the Mikulski Archive for Space Telescopes (MAST): GO-9477, PI: Madore \citep{Madore_2002hst..prop.9477M}; GO 10399, PI: Greenhill \citep{Greenhill_2004hst..prop10399G}; GO-16198, PI: Riess \citep{Riess_2020hst..prop16198R}; GO-16688, PI: Anderson \citep{Anderson_2021hst..prop16688A}; GO-16743, PI: Hoyt \citep{Hoyt_2021hst..prop16743H}. We drizzle together overlapping fields that were taken in the same set of filters ($F814W$ and $F606W$ or $F814W$ and $F555W$) and perform photometry using DOLPHOT \citep{Dolphin_2016ascl.soft08013D, Dolphin_2002MNRAS.332...91D} and the same pipeline used in \cite{Anand_EDDvsCCHP_2021arXiv210800007A}. We apply the same set of DOLPHOT photometric quality cuts (crowd, sharp, object type, error flag, and S/N) for NGC 4258 from \cite{Anand_EDDvsCCHP_2021arXiv210800007A}, which are based on \cite{McQuinn_2017AJ....154...51M}, with the exception of the SNR $F606W$ quality cut, which we set to S/N = 3 instead of S/N = 2.  In field 6 we note excessive error flagging by DOLPHOT of bright stars using the ``benign'' flag=2 (a flag which according to the DOLPHOT manual is usable) which was not common in other fields.  Inspection of these stars indicated no issues we could identify for these stars.  Rather than exclude them, we tested all other field by measuring the difference in the TRGB for all other fields between error cutting and retaining error flag=2 and found no significant difference.  Indeed, the GHOST analyses \citep{Wu_2022arXiv221106354W} use all flag=2 data.  Therefore we concluded it was safe to retain the flag=2 photometry for the field.

\begin{figure*}[ht!]
\epsscale{1}
\plotone{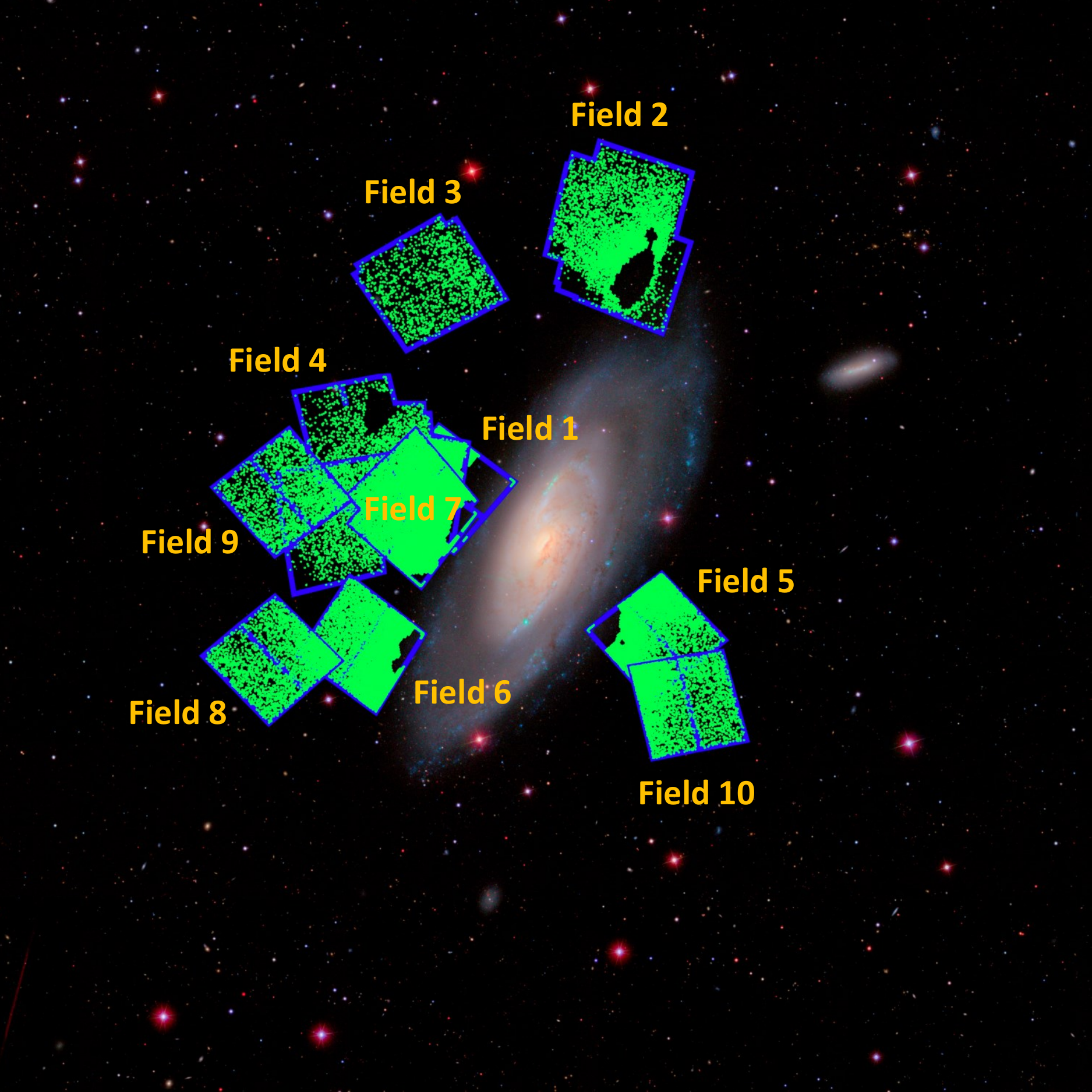}
\caption {Footprints for the 10 fields (blue) and locations of stars that made it past the CATs spatial clipping algorithm (green). Field numbers are labelled in orange. The background image was created by combining $gri$ images from the Sloan Digital Sky Survey \citep{SDSS_DR12_2015ApJS..219...12A}. In this image, north and east correspond to the upward and leftward directions, respectively.}
\label{fig:Field_Locations}
\end{figure*}

We find that the individual fields in Field 4 are clustered in three groups and can be better aligned if drizzled separately. We separate Field 4 into Group 1, Group 2, and Group 3, which we refer to as G1, G2, and G3, respectively. We show these groupings in Appendix \ref{sec:Field4_Groups} and ultimately do not use G3 or the bottom most field in our analysis because the stars are too sparse for us to find a proper alignment solution within DOLPHOT. The locations of the fields can be seen in Fig. \ref{fig:Field_Locations}.

\begin{figure*}
  \centering
  \begin{tabular}{ccc}
    \includegraphics[width=0.3\linewidth]{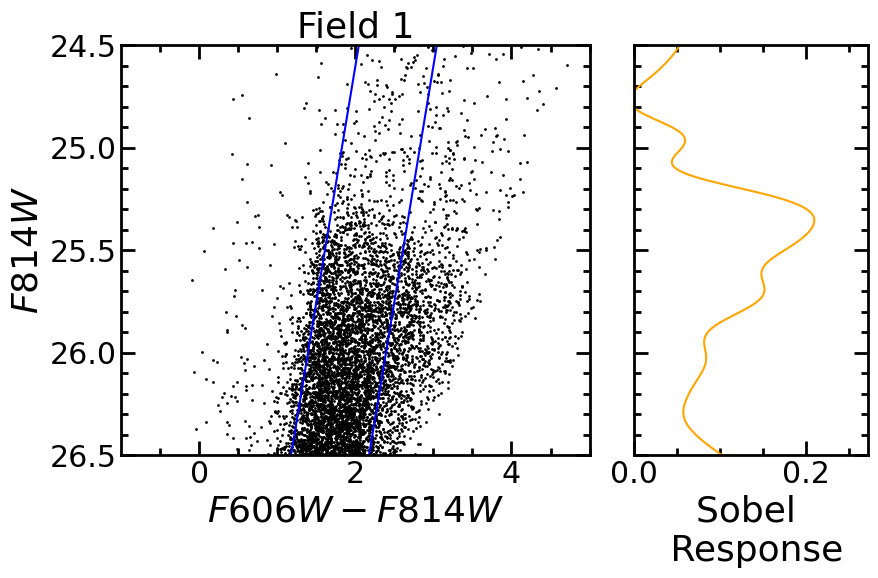} &
    \includegraphics[width=0.3\linewidth]{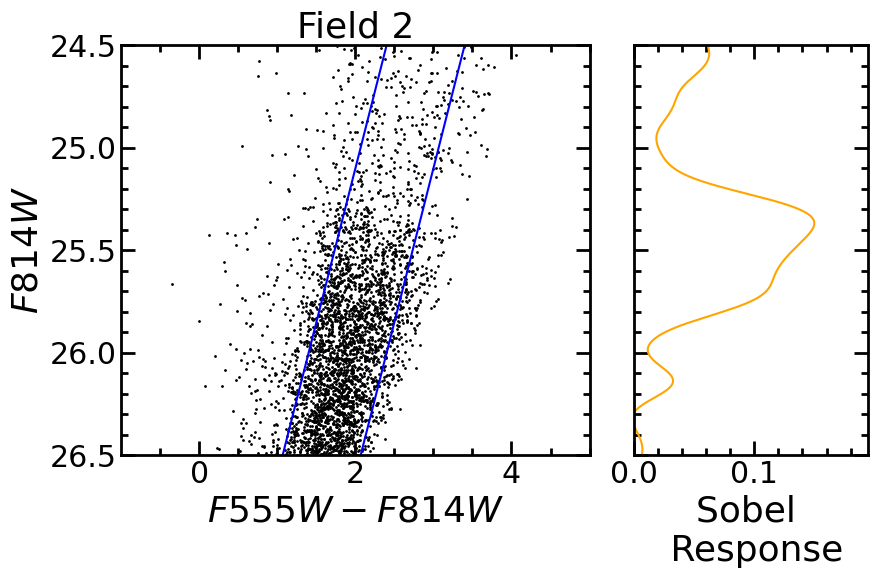} &
    \includegraphics[width=0.3\linewidth]{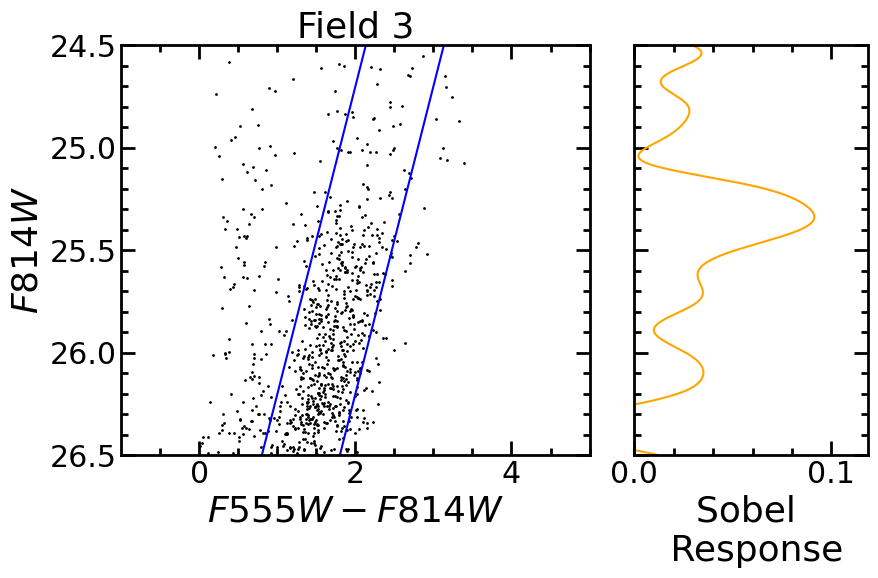} \\
    \includegraphics[width=0.3\linewidth]{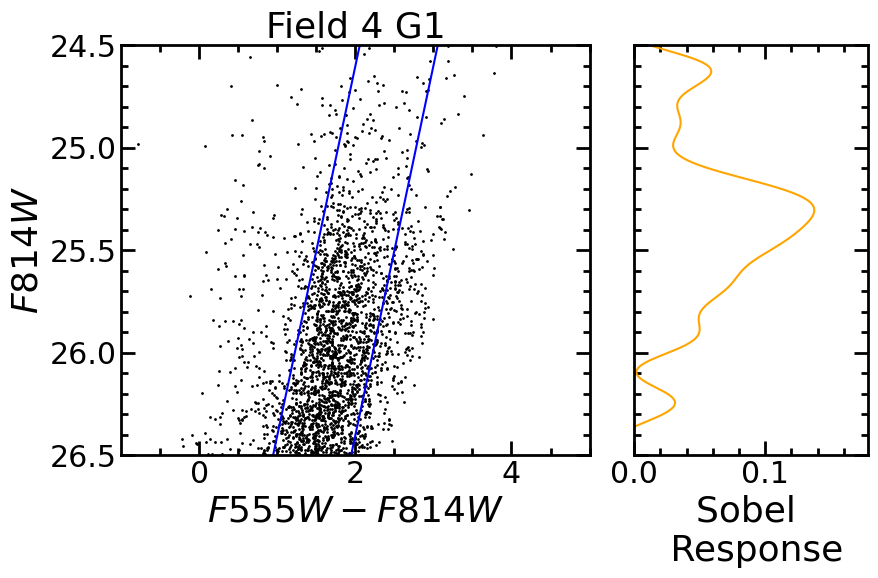} &
    \includegraphics[width=0.3\linewidth]{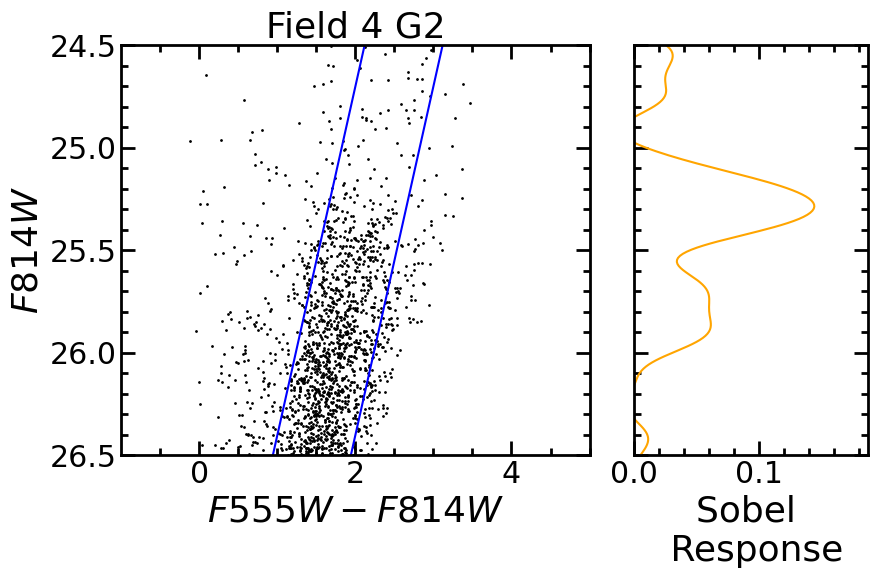} &
    \includegraphics[width=0.3\linewidth]{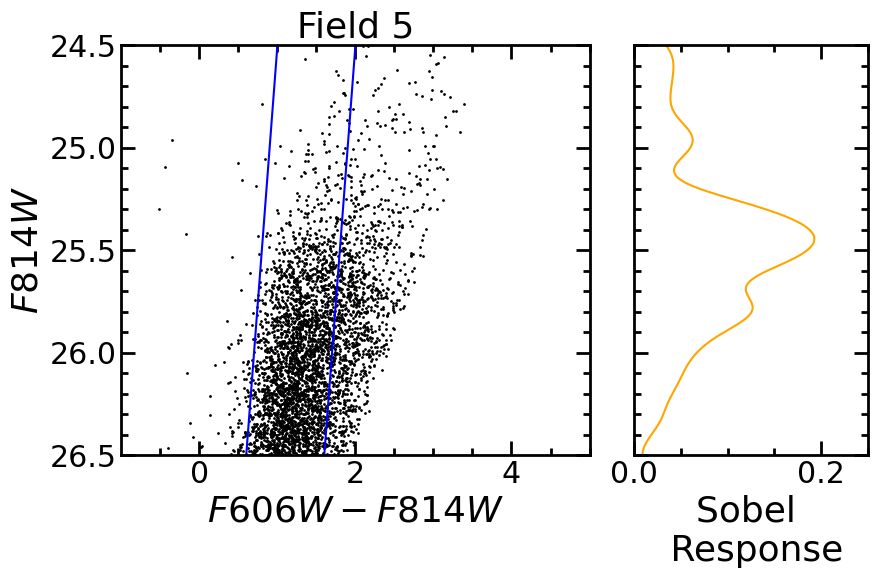} \\
    \includegraphics[width=0.3\linewidth]{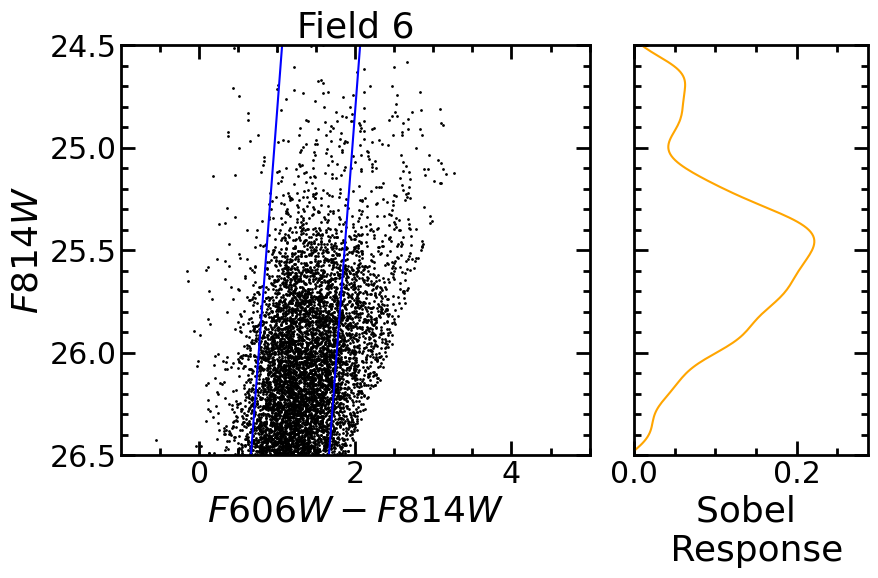} &
    \includegraphics[width=0.3\linewidth]{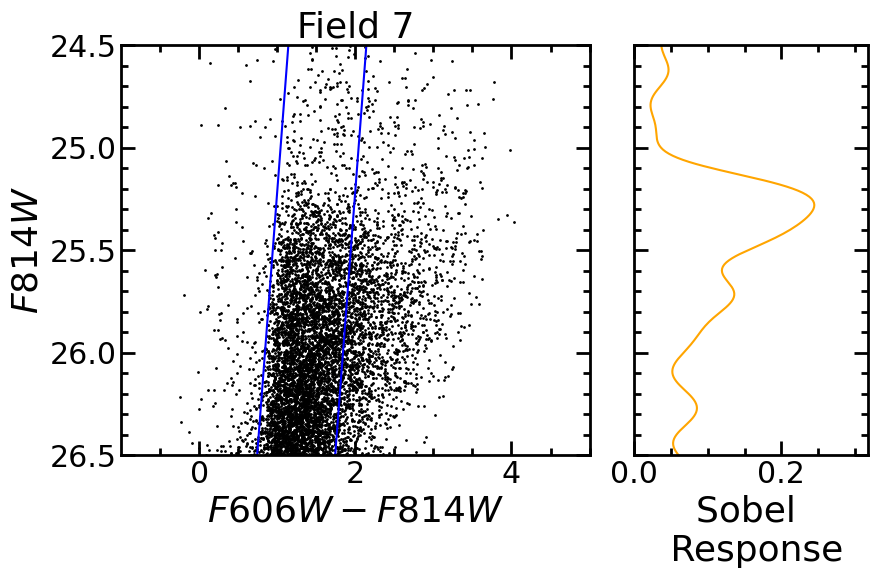} &
    \includegraphics[width=0.3\linewidth]{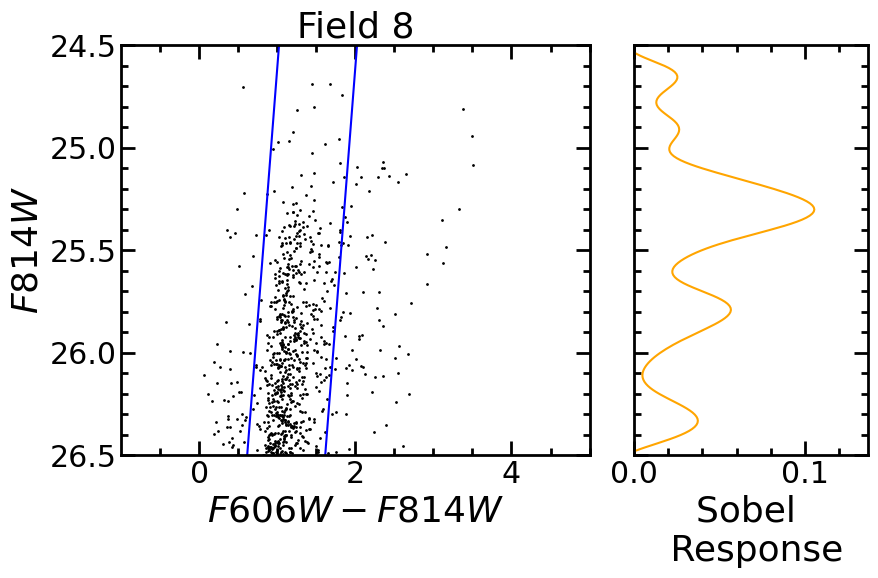} \\
    \includegraphics[width=0.3\linewidth]{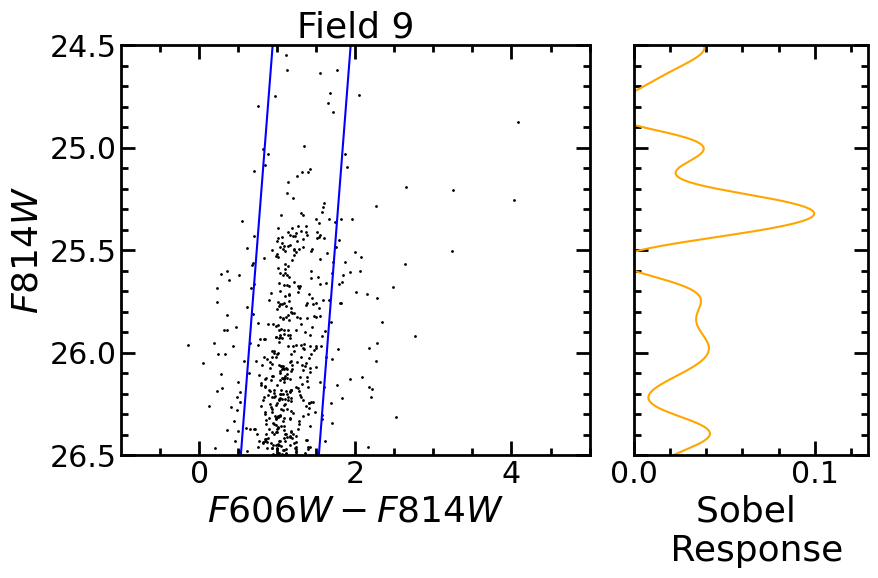} &
    \includegraphics[width=0.3\linewidth]{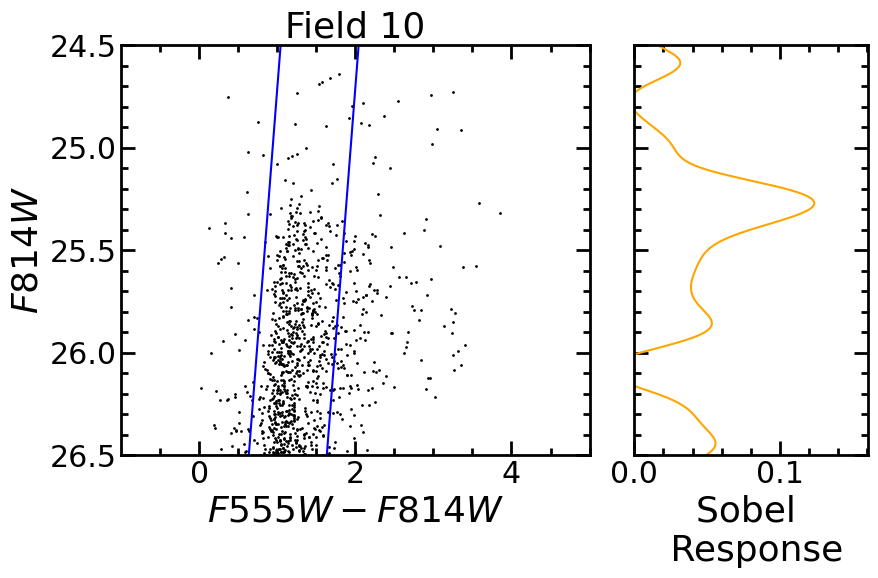} \\
  \end{tabular}
  \caption{Color magnitude diagrams after applying quality and spatial cuts for the fields analyzed in this study are shown in the left panels of each subplot. Internal and external extinction corrections have not yet been applied in this plot. Diagonal blue lines in the left panels for each subplot indicate the optimized color band determined with the CATs algorithm, and the orange lines in the right panels for each subplot show the Sobel response generated with the CATs algorithm with a smoothing parameter of 0.1 and stars within the blue color bands.}
\label{fig:CMDs}
\end{figure*}

We provide a summary of the fields used in Table~\ref{tab:Fields}, and the photometry, along with other intermediate data products, used for this study in a publicly available GitHub repository\footnote{\url{https://github.com/JiaxiWu1018/CATS-H0}}. 

\begin{figure*}[ht!]
\epsscale{1}
\plotone{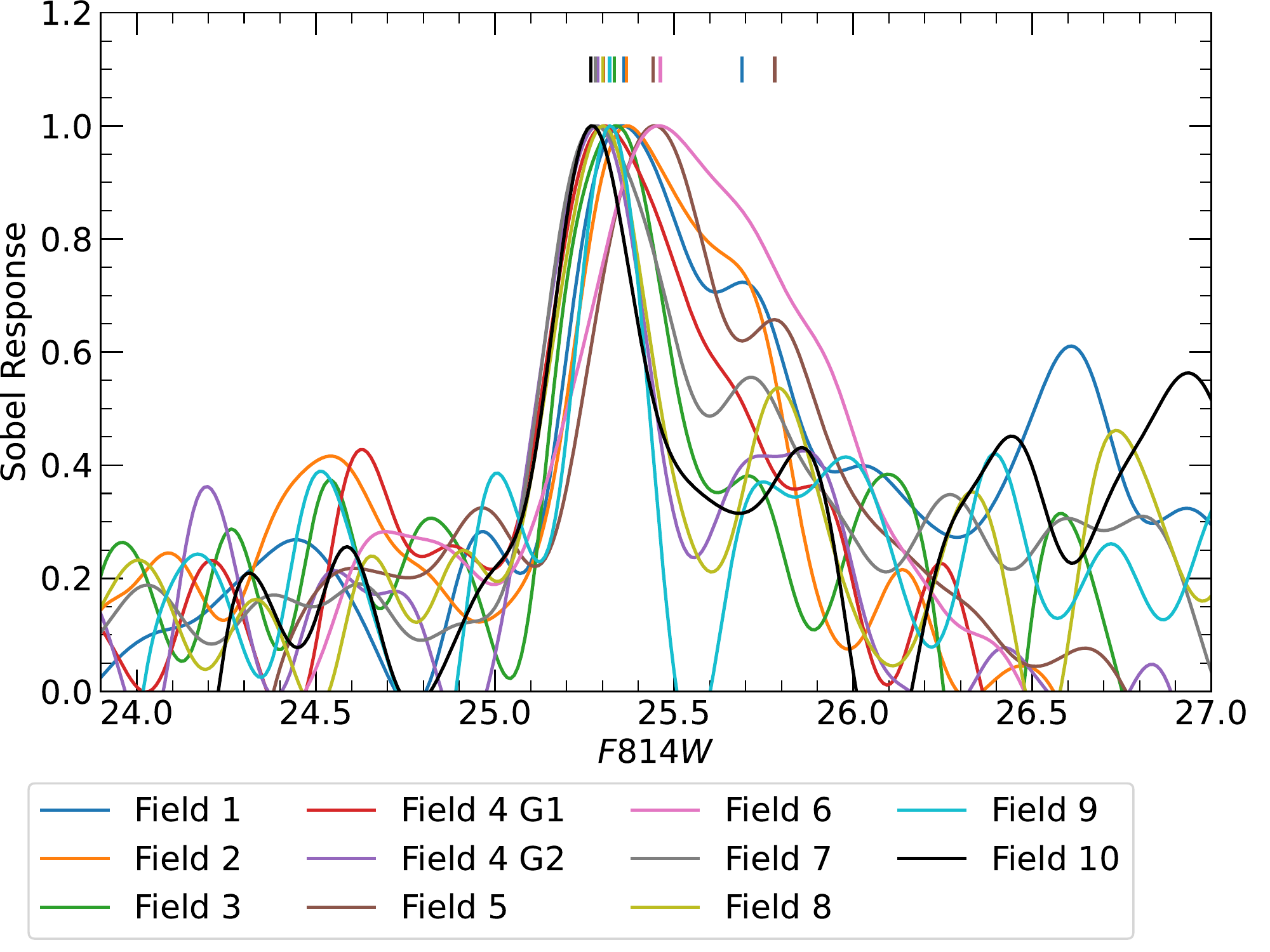}
\caption 
{Normalized Sobel responses for all fields described in Section \ref{sec:Data_Selection} overlayed for comparison. The plotted lines here are identical to the orange lines in Fig. \ref{fig:CMDs} and produced using the CATs algorithm. We measure these Sobel responses for each field using the CATs algorithm by first applying spatial clipping, optimizing and a color band selection, applying 0.1 smoothing to the luminosity function, and evaluate the Sobel response using the same procedure from \cite{Wu_2022arXiv221106354W}. The Sobel responses shown here have not yet been corrected for internal or external extinction. We mark the locations of the measured TRGB from these Sobel responses with tick marks colored with their corresponding Sobel response curve at the top of the figure.}
\label{fig:Combined_Sobel_Responses}
\end{figure*}

\begin{deluxetable*}{cccccc}
\tablecaption{}
\label{tab:Fields}
\tablehead{\colhead{Field Number} & \colhead{Program} & \colhead{PI} & \colhead{Filters} & \multicolumn{2}{c}{Exposure Time (s)}  \\
&&&& $F814W$ & $F555W$/$F606W$ }
\startdata
1 & GO-9477 & Madore & $F814W$/$F555W$ & 2600 & 5700 \\
2 & GO-10399 & Greenhill & $F814W$/$F555W$ & 1040 & 1300  \\
3 & GO-10399 & Greenhill & $F814W$/$F555W$ & 1040 & 1300 \\
4 & GO-10399 & Greenhill & $F814W$/$F555W$ & 1040 & 1300 \\
5 & GO-16198 & Riess & $F814W$/$F606W$ & 640 & 785  \\
6 & GO-16198 & Riess & $F814W$/$F606W$ & 640 & 785  \\
7 & GO-16688 & Anderson & $F814W$/$F606W$ & 2250 & 2250  \\
8 & GO-16743 & Hoyt & $F814W$/$F606W$ & 5146 & 5146 \\
9 & GO-16743 & Hoyt & $F814W$/$F606W$ & 2614 & 5146 \\
10 & GO-16743 & Hoyt & $F814W$/$F606W$ & 5146 & 5146 \\
\enddata
\caption{Summary table for the fields used in this study. Columns from left to right: field number, program numbers, principal investigators, filter sets, and exposure times for each filter.}
\end{deluxetable*}

\section{Tip Properties in NGC 4258}
\label{sec:Multi-field_Average}

It is useful to first examine the direct tip measurements and properties before applying any additional corrections. To measure the tip, we use the same algorithm from \cite{Wu_2022arXiv221106354W}. This algorithm applies spatial clipping to remove contamination from young stars, optimizes a color band, smooths the luminosity function within the optimized color band with Gaussian-windowed, Locally Weighted Scatterplot Smoothing (GLOESS) with a smoothing parameter of $s$ = 0.1, applies a weighted Sobel filter to obtain edge-detection responses (EDRs), and finds the tip based on local inflection points along the luminosity function. The detected tips and EDRs using a smoothing parameter of $s$ = 0.1 near the RGB among the 11 different fields around NGC 4258 are shown in Fig. \ref{fig:CMDs} and compared in Fig. \ref{fig:Combined_Sobel_Responses}.   They appear quite inhomogeneous even after we reject tips with contrast ratios less than 2; we find that these latter detections have high noise due to a weakly defined break and in many cases correspond to the tip of the AGB rather than red giant branch (RGB).

The range of detected tips is several tenths of a magnitude.  The average of the 11 field tips is $F814W_0=25.37$~mag with a dispersion of 0.099 mag.  This value uses the external extinction estimate of 0.030~mag from \cite{Schlafly_2011ApJ...737..103S} and $A_{F814W}$/E(B-V) = 2.107 (R. Anderson, private communication), consistent with \cite{Anderson_2023arXiv230304790A}, but is not corrected for internal extinction \citep{Anderson_2022A&A...658A.148A}, which has an expected average $A_{F814W}=0.013$ mag using \cite{Menard_2010MNRAS.405.1025M} \cite[see][for a discussion]{Wu_2022arXiv221106354W}. For two cases, fields 1 and 5, we computed their value by averaging rather than selecting one of two tips.

If instead we average the fields by the number of stars below the tip each contains (hence weighted by the inverse Poisson error as though together they were one big field) we find $F814W_0=25.42$~mag, where we present this average in the context that a change in methodology can significantly affect the result.  It is important to note here that the field dispersion of $\sigma=0.10$ mag dominates the statistical bootstrap precision of $<$ 0.03 mag per field. This dispersion cannot be attributed to disk contamination due to our use of spatial masking or the small number of younger stars that remain after spatial clipping in the corners of the fields (for instance, with fields 1 and 5).    We also note that none of these fields are outliers (more than 2 $\sigma$ from the mean, full range of 25.27 to 25.61 mag). Because of this field-to-field variation, the simple mean tip of NGC 4258 is no more accurate than $\sigma \sim 0.03$ mag.  We also note that our results are in good agreement with the tip measurements in different fields used by \cite{Jang_2021ApJ...906..125J} and \cite{Anand_EDDvsCCHP_2021arXiv210800007A}.  

With the maser distance, the simple means result in $M_{F814W,0}=-4.05 \pm 0.03$ mag, or $M_{F814W,0}=-4.00 \pm 0.03$ weighted by star number, which serves as a useful reference for studies of the metal-poor tip which do not embark on further standardization. We provide the measured tips, uncertainties, contrast ratios, TCR corrected tips, internal extinction, and foreground extinction values in Table \ref{tab:TCR}. We also explore more tip measurement properties in Appendix \ref{sec:Measuring_Tips}.

\section{Tip-Contrast Ratio Relationship}
\label{sec:TCR}

It is not entirely clear what produces the large variance of EDRs shown in the fields of NGC 4258, and changing the level of smoothing changes the magnitude of the variance\ \cite[see also][]{Anderson_2023arXiv230304790A}. However, to reduce a bias due to a mismatch when they are used to calibrate the TRGB in other fields (e.g., in SN Ia hosts), we seek to standardize or homogenize the tip measurements across different fields.  Color has been used \citep{ Rizzi_2007ApJ...661..815R, Jang_2017ApJ...835...28J} for fields where the TRGB has been measured over a wide color range.  However, the variance we see is present after selecting a narrow color range ($\Delta$ $V - I$ $\sim$ 1.0).  Here we follow the approach by \cite{Wu_2022arXiv221106354W} who found an empirical relation between the observed tip magnitude of a field and the measured contrast-ratio (the ratio of stars above versus below the tip, parameterized as $R$).  

\begin{figure*}[ht!]
\epsscale{1}
\plotone{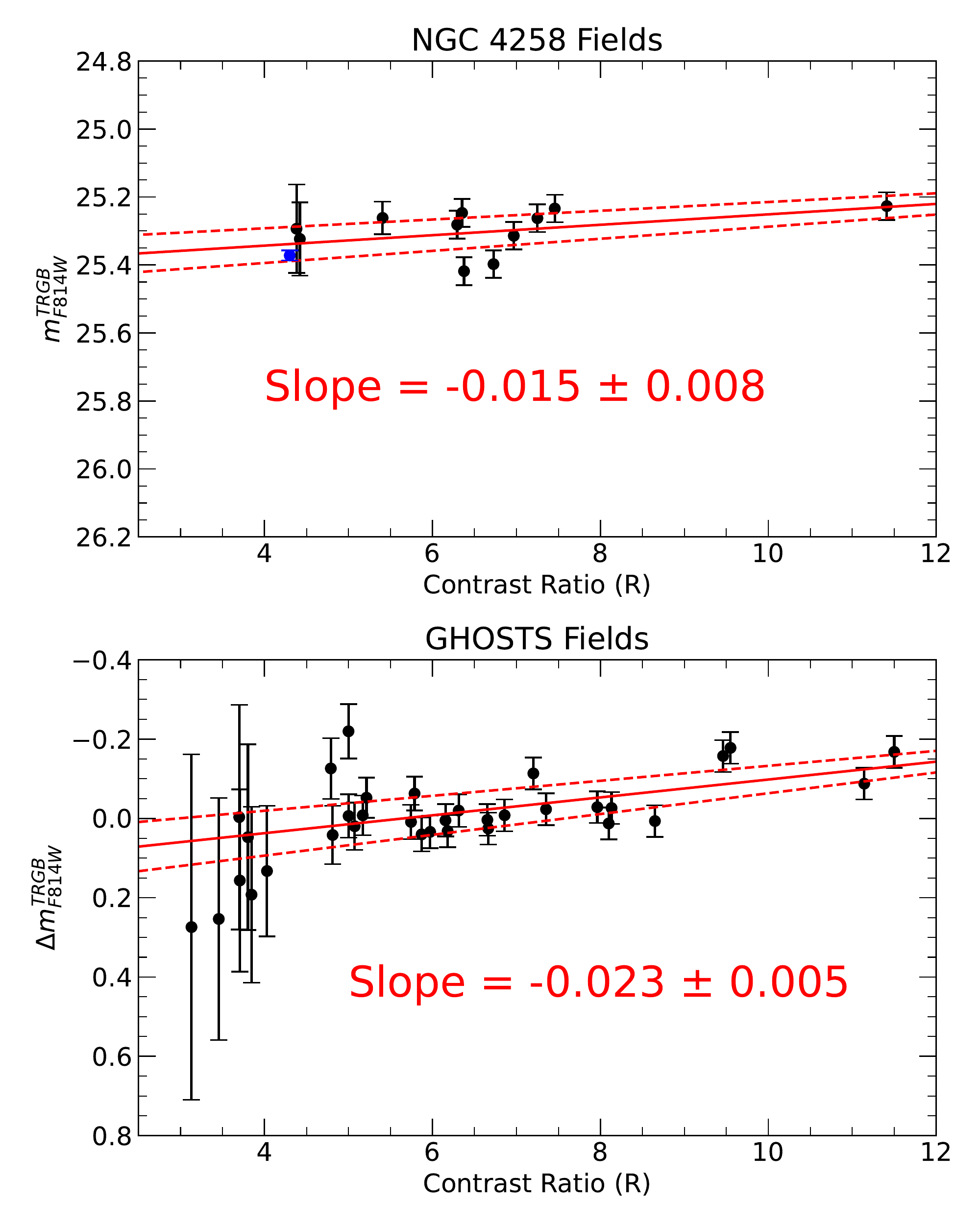}
\caption 
{Measured tips as a function of contrast ratios using the CATs algorithm described in \cite{Wu_2022arXiv221106354W} using the NGC 4258 fields from Fig. \ref{fig:Field_Locations}, Fig. and Table \ref{tab:Fields} (top) and GHOSTS fields \citep{Radburn-Smith_2011ApJS..195...18R} (bottom). An approximated tip-contrast ratio pair from Fig. 10 in \cite{Jang_2021ApJ...906..125J} is added in the top panel in blue for comparison purposes only. We note that their measurement setup would likely produce a TCR different from the one calibrated here.}
\label{fig:TCR}
\end{figure*}

%The wide range of detected tips 
We characterize and employ the empirically determined relation between the brightness of the tip and the contrast ratio. We measure the tips and contrast ratios for the 11 fields described in Section \ref{sec:Data_Selection} and present them in Table \ref{tab:TCR}. For fields with more than one tip, we average the tips measured with CATs in each field (weighted by their contrast ratios), correct for internal and foreground extinction, and plot these values in the top panel of Fig.~\ref{fig:TCR}. Tip uncertainties were estimated using the model determined in \cite{Wu_2022arXiv221106354W}:

\begin{equation}
 %\sigma=\sqrt{\left[\left({\frac{0.4}{R-2}}\right)\left(\frac{200}{N}\right)^{0.3}\right]^2+0.03^2}\;\mathrm{mag}
 \sigma=\sqrt{\left[\left(\frac{2e^{1.5(3-R)}}{e^{1.5(3-R)}+1}\right)\left(\frac{1}{\Nminusone-100}\right)^{0.1}\right]^2+0.04^2}\;\mathrm{mag}
 \label{eqn:error}
\end{equation}

We look for a relationship of the form 

\begin{equation}
\label{eq:TCR_General}
    m^{R = 4}_{TRGB} = m_{TRGB} - (R - 4.0) \times m
\end{equation}

\noindent where $m^{R = 4}_{TRGB}$ is the fiducial tip at a contrast ratio of 4, $m_{TRGB}$ is the measured tip before correction, $R$ is the contrast ratio, and $m$ is the slope of the tip-contrast ratio relationship. We first determine $m$ via a linear least squares regression to the measured tips in Table \ref{tab:TCR}, and find $m = -0.015 \pm 0.008$~mag/R. We do not find any outliers outside a 3 $\sigma$ clip.  This value is 1 $\sigma$ from the calibration by \cite{Wu_2022arXiv221106354W} of $-0.023 \pm 0.005$~mag per unit contrast ratio. A weighted average of these two measurements yields $-0.021 \pm 0.004$~mag/R. Following Eq.~\ref{eq:TCR_General}, we then measure a tip at a contrast ratio of 4 of $m^{R = 4}_{F814W} = 25.361 \pm 0.0136$~mag. We subtract the maser distance of $\mu_{N4258} = 29.397 \pm 0.0324$~mag  from \cite{Pesce_2020ApJ...891L...1P} to find a zero-point of 

\begin{equation}
   M^{R = 4}_{F814W} = -4.036 \pm 0.035 \textnormal{ mag} - (R - 4) \times 0.021
\end{equation}

Removing the internal extinction correction gives $m^{R = 4}_{F814W} = 25.372 \pm 0.0136$~mag and a zero-point of

\begin{equation}
    M^{R = 4}_{F814W} = -4.025 \pm 0.035 \textnormal{ mag} - (R - 4) \times 0.021
\end{equation}
We summarize the mean and standardized tips across all fields from Section \ref{sec:Multi-field_Average} and this section in Table \ref{tab:tip_Summary}. 

\begin{deluxetable*}{cccccccc}
\tablecaption{}
\label{tab:TCR}
\tablehead{\colhead{Field} & \colhead{$m_{F814W, TRGB}^{R = 4}$}  &\colhead{$\sigma_{TRGB}*$}  & 
\colhead{R} &\colhead{$\Nminusone$} &\colhead{$m^{TRGB}_{F814W}$} &\colhead{$A_{I}(Int.)$} &\colhead{$A_I(MW)$}}
\startdata
1 & 25.375 & 0.041 & 7.00 & 3193 & 25.360 & 0.019 & 0.03 \\
1 & 25.619 & 0.462 & 2.91 & 4529 &25.690 & 0.019 & 0.03 \\
2 & 25.332 & 0.108 & 4.42 & 1971 & 25.367 & 0.014 & 0.03 \\
3 & 25.301 & 0.130 & 4.38 & 480 & 25.334 & 0.011 & 0.03 \\
4 G1 & 25.291 & 0.048 & 5.41 & 1386 & 25.304 & 0.013 & 0.03 \\
4 G2 & 25.296 & 0.041 & 6.35 & 902 & 25.288 & 0.012 & 0.03 \\
5 & 25.453 & 0.041 & 6.75 & 2451 & 25.441 & 0.016 & 0.03 \\
5 & 25.707 & 0.567 & 2.64 & 3199 & 25.781 & 0.016 & 0.03 \\
6 & 25.468 & 0.041 & 6.38 & 4120 & 25.462 & 0.014 & 0.03 \\
7 & 25.306 & 0.041 & 7.46 & 3550 & 25.280 & 0.016 & 0.03 \\
8 & 25.331 & 0.041 & 7.25 & 474 & 25.302 & 0.010 & 0.03 \\
9 & 25.330 & 0.041 & 6.29 & 302 & 25.320 & 0.009 & 0.03 \\
10 & 25.382 & 0.040 & 11.41 & 531 & 25.268 & 0.011 & 0.03 \\
\enddata
\caption{* From Equation \ref{eqn:error}. From left to right: field number, tip corrected to a fiducial contrast ratio of R = 4, tip errors, contrast ratios, number of stars 1 mag fainter than the measured tip, original tips before TCR correction, internal extinction estimated with \cite{Menard_2010MNRAS.405.1025M}, and Milky Way extinctions estimated with \cite{Schlafly_2011ApJ...737..103S}. Tip, error, and contrast ratios were measured using the unsupervised CATs algorithm. The tips shown in the fifth column from the left have not been corrected for internal and foreground extinction.}
\end{deluxetable*}

\begin{deluxetable*}{ccccc}
\tablecaption{}
\label{tab:tip_Summary}
\tablehead{\colhead{Method} & \colhead{$m^{TRGB}_{F814W}$}  & \colhead{MW $A_{F814W}$} & \colhead{Int. $A_{F814W}$} &\colhead{TCR}}
\startdata
Average (by Field) & 25.37 & 0.030 & Not Applied & None  \\
Average (by Stars) & 25.42 & 0.030 & Not Applied & None  \\ 
Standardization & $25.372 \pm 0.0136$ & 0.030 & Not Applied & $-0.021 \pm 0.004$\\
Standardization & $25.361 \pm 0.0136$ & 0.030 & 0.011 & $-0.021 \pm 0.004$\\
\enddata
\caption{Summary of tip measurements from Sections \ref{sec:Multi-field_Average} and \ref{sec:TCR}. The standardized tip is taken to be at a fiducial contrast ratio of $R = 4$. For the tip measurements shown here, we applied foreground extinction correction but not internal extinction to facilitate better comparison to past studies in the literature.}
\end{deluxetable*}

\subsection{\ion{H}{1} Regions}
\label{sec:tip_Summary}

\begin{figure*}[ht!]
\includegraphics[width=.48\linewidth]{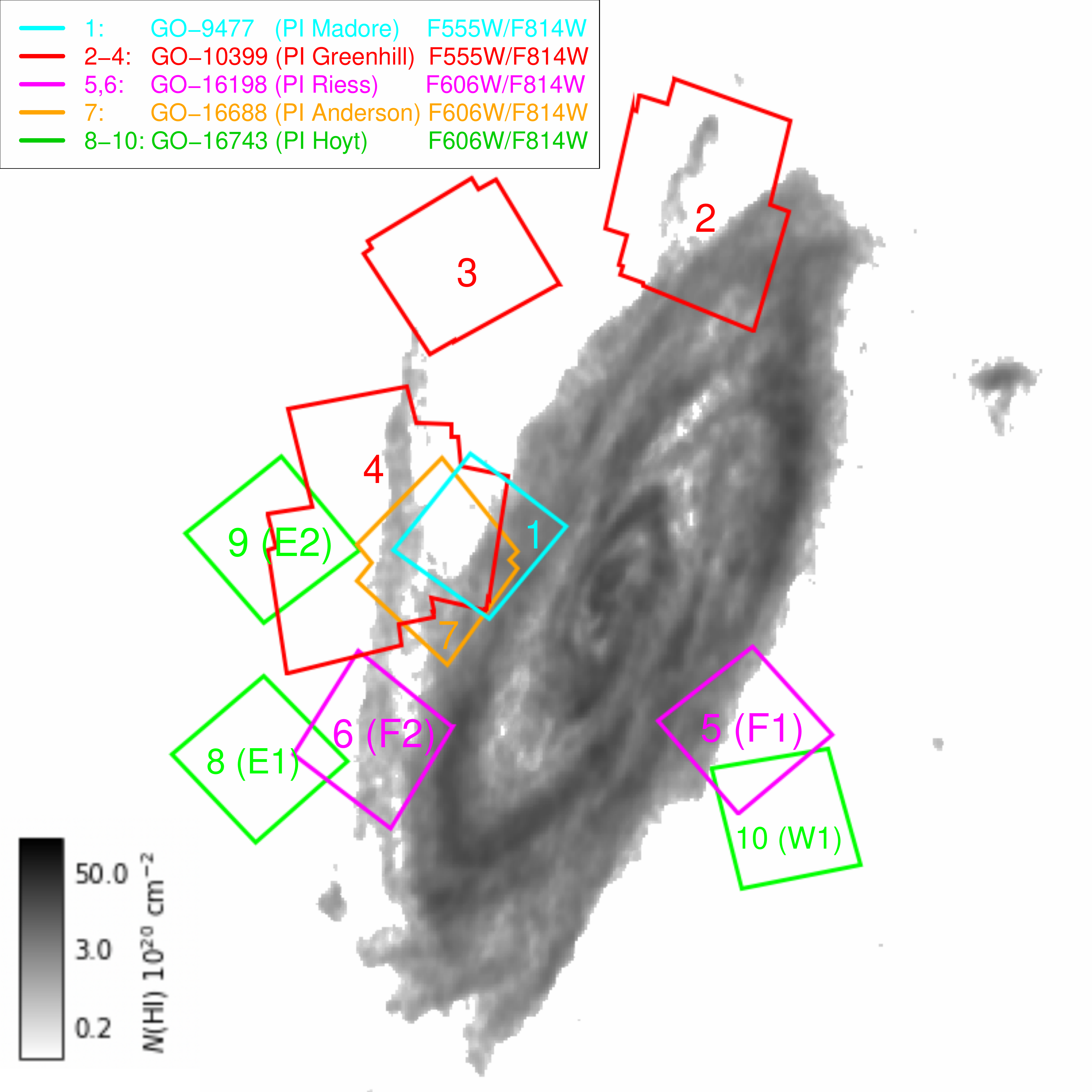}%
%\label{subfig:a}%
\hfill
\includegraphics[width=.48\linewidth]{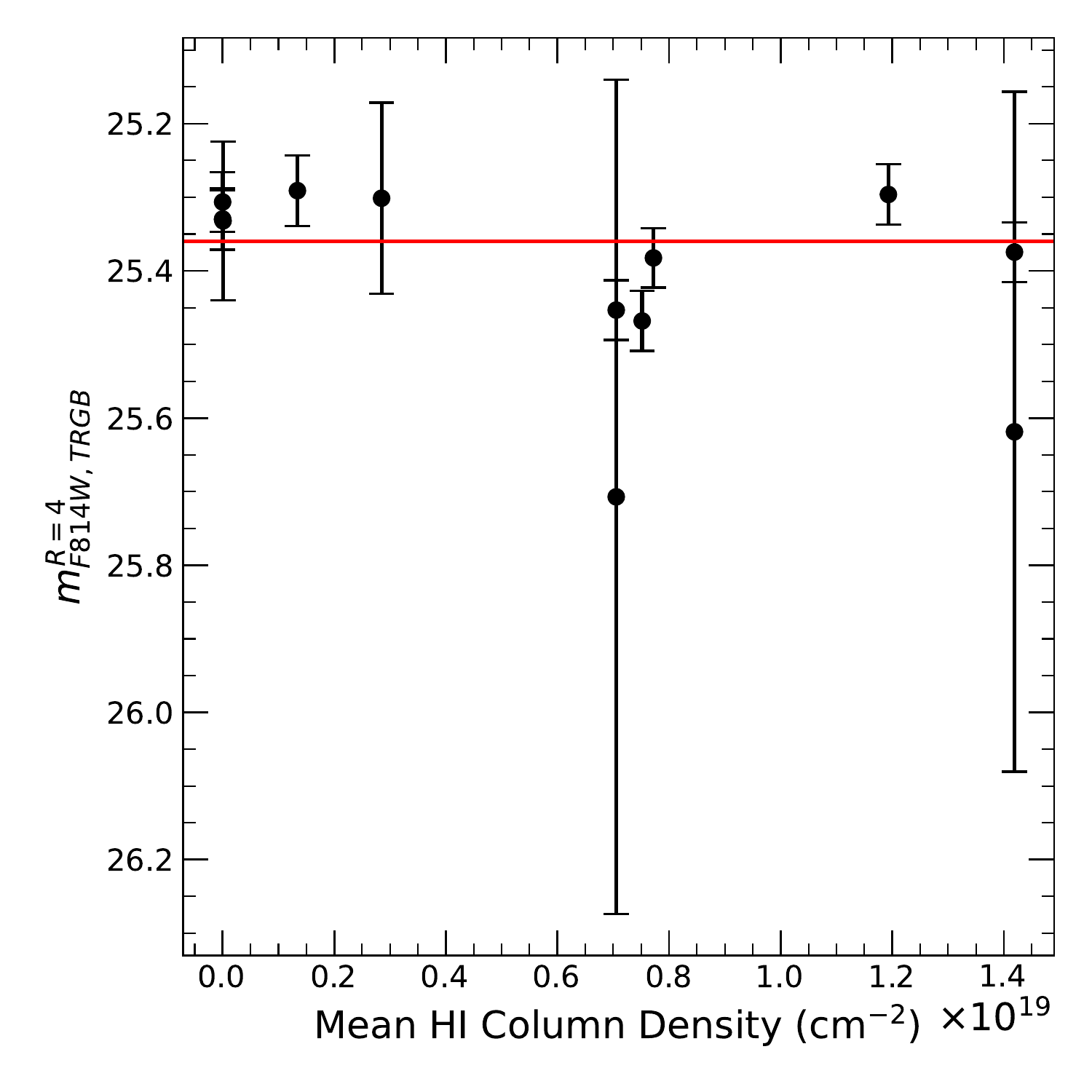}
\caption {\textit{Left}: Locations of the 10 fields analysed in this study overplotted on a \ion{H}{1} density map from \cite{Heald_2011A&A...526A.118H}. \textit{Right}: TCR corrected tips from Table \ref{tab:TCR} as a function of the mean \ion{H}{1} column density. We do not find a significant trend between the corrected tip and \ion{H}{1} column density.}
\label{fig:HI}
\end{figure*}

We examine whether the tip measurements correlate with the amount of \ion{H}{1} present in each field. We obtain \ion{H}{1} column density measurements from \cite{Heald_2011A&A...526A.118H}. In the left panel of Fig.~\ref{fig:HI}, we overlay the field contours on the \ion{H}{1} map from \cite{Heald_2011A&A...526A.118H}. In the right panel of Fig.~\ref{fig:HI}, we plot the TCR-corrected tips from Table~\ref{tab:TCR} as a function of the mean \ion{H}{1} column density for each field using the \ion{H}{1} map. We plot a horizontal red line at 25.37~mag for reference, which corresponds to the TCR corrected tip from Section \ref{sec:TCR}. We fit the points in Fig.~\ref{fig:HI} with a linear least squares regression and find a slope of $0.04 \pm 0.03$~mag/cm$^2$. We do not find a significant trend (i.e. $< 1.5 \sigma$) between \ion{H}{1} column density and measured tip after applying the CATs algorithm and TCR correction. If we add back in stars that were removed with spatial clipping, we find a fit of $0.05 \pm 0.03$~mag/cm$^2$.

\section{TCR Simulations}
\label{sec:ArtpopTCR}

In this section, we investigate whether a tip-contrast ratio relationship can be created using the CATs algorithm applied to simulated stellar populations. To do this, we use ArtPop, which is a Python package that can generate synthetic stellar populations and artificial images of stellar systems \citep{ArtPop_2022ApJ...941...26G} using Modules for Experiments in Stellar Astrophysics \citep[MESA;][]{Paxton_2011ApJS..192....3P, Paxton_2013ApJS..208....4P, Paxton_2015ApJS..220...15P} Isochrones and Stellar Tracts (MIST) \citep{Dotter_2016ApJS..222....8D, Choi_2016ApJ...823..102C}. While these simulations include noise due to interpolation errors and choices of stellar models, we attempt to reproduce a TCR in general within the limitations of the single age and metallicity synthetic populations using informed assumptions about the underlying populations. We do not necessarily attempt to recreate a population of stars in a halo field.

\begin{figure*}[ht!]
\epsscale{1}
\plotone{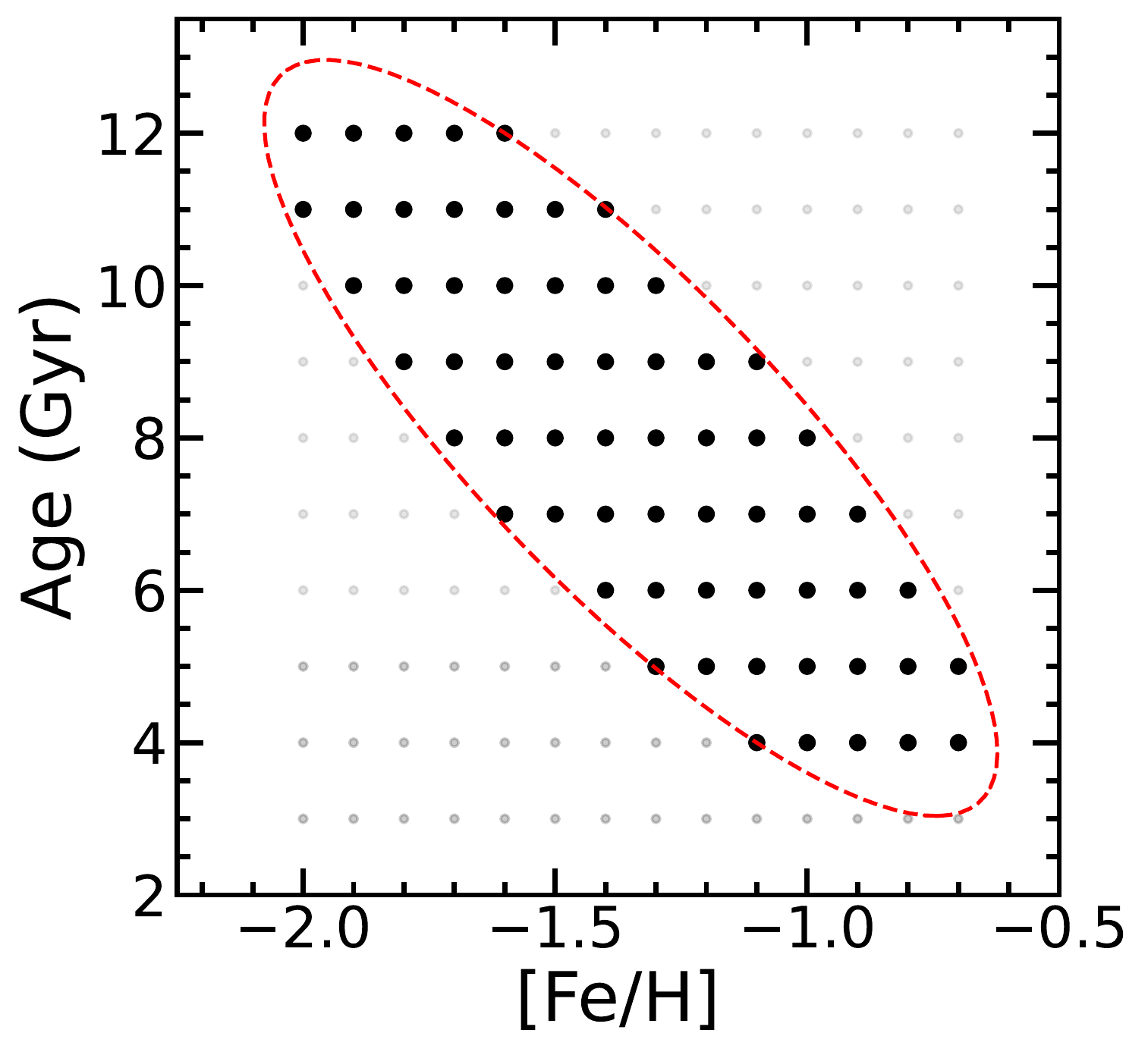}%
%\label{subfig:a}%
\caption {Grid of ages (3 - 12 Gyr) and metallicities ([Fe/H]; -2 to -0.7) used to simulate stellar populations for Fig. \ref{fig:Sim_Noise_Smoothing_Grid}. To approximate an old, metal-poor to young, metal-rich trend, we draw an ellipse along this axis on the plot. Points that fall inside this ellipse are shown as large, solid dots while points that lie outside this ellipse are shown as smaller fainter dots.}
\label{fig:Sim_Noise_Smoothing_Grid_Ellipse}
\end{figure*}

We simulate a grid of stellar populations of single ages and metallicities from 3 to 12 Gyr and [Fe/H] from $-2$ to $-0.7$ (see Fig. \ref{fig:Sim_Noise_Smoothing_Grid_Ellipse}). The total mass for each population is 6 billion solar masses. To better reflect actual measurement conditions, we add Gaussian distributed noise following an exponential model of $\sigma = Ae^{Bm}$ where $\sigma$ is the noise and $m$ is the magnitude of the star. We calculate the parameters $A$ and $B$ by fitting the uncertainties estimated using artificial stars for NGC 1448 from the Extragalactic Distance Database \citep[EDD;][]{Tully_2009AJ....138..323T, Anand_EDD_2021AJ....162...80A}. We selected NGC 1448 as a representative galaxy to use for galaxies used to measure $H_0$ in a companion study \citep{Scolnic_2023arXiv230406693S}. For $\sigma_{F606W}$ and $\sigma_{F814W}$, we find $A = 1$ and $B = 0.710$, and $A = 1.684$ and $B = 0.721$, respectively.  Because NGC 1448 is at a different distance than NGC 4258, we apply an additional scaling so that the errors are 0.05~mag at $F814W = -4$~mag and $F606W = -3$~mag, which we take to be approximately the location of the tip. Then, we use the CATs algorithm with the baseline parameters described in Section \ref{sec:TCR} to measure the tip and contrast ratio for each simulated population of stars.

\begin{figure*}[ht!]
\epsscale{1}
\plotone{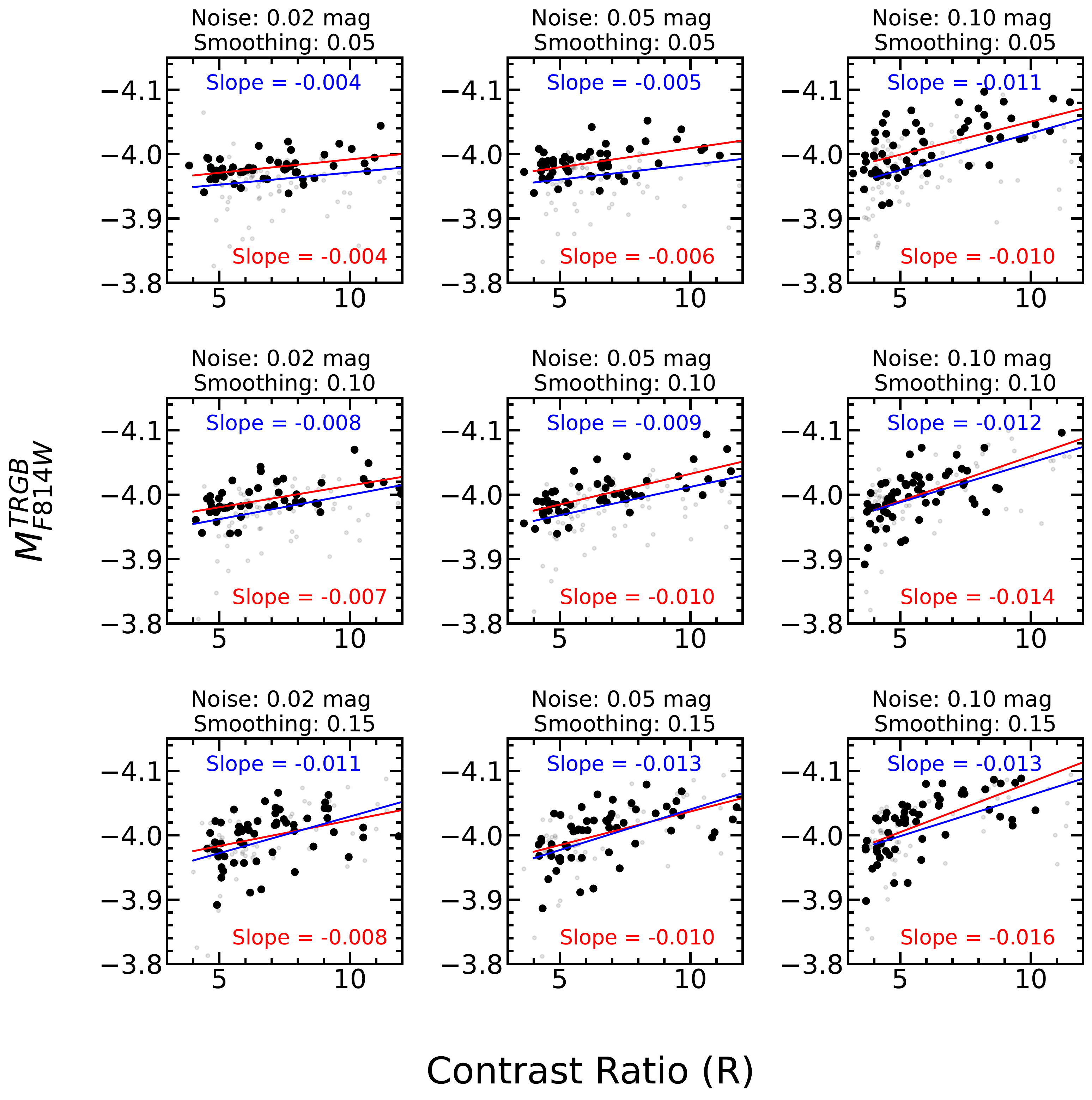}
\caption {Tip measurements as a function of contrast ratios using the CATs algorithm on simulated populations of single ages (3 - 12 Gyr) and metallicities ([Fe/H]; -2 to -0.7) generated with ArtPop with the selection shown in Fig. \ref{fig:Sim_Noise_Smoothing_Grid_Ellipse}. We then apply the CATs algorithm on this grid of simulated populations with a grid of noise (0.02, 0.05, and 0.10~mag) and smoothing values (0.05, 0.10, 0.15). We restrict and fit each subplot to contrast ratios between 3 and 12. We also restrict the tips to be fainter than $-4.1$~mag as we find tips brighter than this often correspond to the tip of the AGB rather than TRGB. We show the points that have ages and metallicities that lie inside the ellipse from Fig. \ref{fig:Sim_Noise_Smoothing_Grid_Ellipse} as large, solid black dots and points with ages and metallicities that lie outside the ellipse as smaller, fainter dots. Linear least squares fits and their slopes are shown in red (large dots) and blue (all dots including large and small). Fits are unweighted.}
\label{fig:Sim_Noise_Smoothing_Grid}
\end{figure*}

We first plot the measured tips and contrast ratios in Fig. \ref{fig:Sim_Noise_Smoothing_Grid}. We limit the range of interest to contrast ratios between 3 and 12 to more closely match the contrast ratios observed in NGC 4258 and for GHOSTS galaxies from \cite{Wu_2022arXiv221106354W}. We note that not all age and metallicity combinations plotted may exist in nature, and real halos are not simple populations. To select a more realistic sample, we draw an ellipse roughly containing populations that follow old and metal-poor to young and more metal-rich. We show this selection in Fig. \ref{fig:Sim_Noise_Smoothing_Grid_Ellipse} and plot points that lie inside the ellipse in both Fig. \ref{fig:Sim_Noise_Smoothing_Grid} and \ref{fig:Sim_Noise_Smoothing_Grid_Ellipse} as large, solid dots. Points that lie outside the ellipse are shown as lighter, smaller dots. For instance, 3 Gyr populations may be too young for stars to reach the tip \citep{Serenelli_2017A&A...606A..33S}, which is why they are excluded form the ellipse.

To measure the slope for the large, solid points, we apply an unweighted linear least squares fit and find a slope of $-0.010 \pm 0.002$~mag per unit R. Similar to empirical calibration found in Section \ref{sec:TCR}, we find a significant inverse relationship between the tip magnitude and contrast ratio (i.e. the tip magnitude becomes brighter with increasing contrast ratio). This fit corresponds to the center subplot in Fig. \ref{fig:Sim_Noise_Smoothing_Grid}.

\subsection{Impact of Smoothing}
\label{sec:Impact_of_Smoothing}

To better understand the impact of the measurement technique itself on the TCR, we repeat the procedure described above with a grid of values for smoothing and noise. We apply 1) GLOESS smoothing values of $s$ = 0.05, 0.10, and 0.15, 2) noise values of 0.02~mag, 0.05~mag, and 0.10~mag to all simulated magnitudes, and 3) the same age and metallicity range as used above. We fit the points with a linear least squares regression and plot the results in Fig. \ref{fig:Sim_Noise_Smoothing_Grid}. We find that in general, the fitted slopes increase in steepness with increasing smoothing and noise, although this trend could vary with different luminosity functions \cite[see][]{Anderson_2023arXiv230304790A}. We notice that the smoothing function with a large smoothing of 0.15 attempts to smooth out the tip discontinuity. This results in a smoothed luminosity function with larger number of stars brighter than the tip when compared to a smoothed luminosity function using a lower smoothing parameter. This occurs despite the bins in the unsmoothed luminosity function remaining the same. We describe this effect in further detail in Appendix \ref{sec:Diff_Smoothing}. 

These populations are generated using homogeneous ages and metallicities, whereas observed fields will likely contain a mixture of different stellar properties. We note that we do not attempt to reproduce the observed halo fields, as creating a simulated population that accurately reflects the possible mixtures of stellar properties involves detailed considerations that are outside the scope of this study. In addition, simulations can also include additional errors due to interpolation and variations in the models, for instance. However, this independent replication of the tip-contrast ratio relationship using simulated populations suggests the tip-contrast ratio relationship is not an artifact of field placement and can be impacted by smoothing and noise.

\section{Discussion}
\label{sec:Discussion}

We calibrated the tip-contrast ratio relationship in NGC 4258 and provide a standardized tip zero-point at a fiducial contrast ratio of 4 that can be used to anchor extragalactic distance measurements. We generated simulated populations using ArtPop \citep{ArtPop_2022ApJ...941...26G} and reproduced a tip-contrast ratio relationship with the CATs algorithm demonstrating that the tip-contrast ratio relation not originate from field placement and is impacted by noise and smoothing. 

We investigated the potential effects of \ion{H}{1} column density on our results. \cite{Jang_2021ApJ...906..125J} and \cite{Anand_EDDvsCCHP_2021arXiv210800007A} argue that \ion{H}{1} regions may affect the measured tip by contaminating the sample with non-negligible internal extinction and younger stars in the halo. However, \cite{Wu_2022arXiv221106354W} argues that the results from \cite{Menard_2010MNRAS.405.1025M} demonstrate that extinction in the halo where the tip is typically measured is negligible, even in the presence of \ion{H}{1} gas. In our analysis, we remove younger stars using the spatial clipping process described in \cite{Wu_2022arXiv221106354W}, which was based on the concepts from \cite{Anand_2018AJ....156..105A, Anand_EDDvsCCHP_2021arXiv210800007A}. 

We do not find evidence of a significant relationship between the measured tip and amount of \ion{H}{1} gas present in the field after spatial clipping. This suggests that the measurement presented above is not affected by the trend observed between the measured tip and \ion{H}{1} column density described in \cite{ Beaton_2018SSRv..214..113B, Jang_2021ApJ...906..125J, Anand_EDDvsCCHP_2021arXiv210800007A}. We note that even if there was a remaining relationship between \ion{H}{1} column density and the tip, the low uncertainty of 0.005 to 0.008 mag per unit contrast ratio in the fitted tip-contrast ratio slope presented in Section \ref{sec:TCR} suggests that any effects from measuring the tip using stars in \ion{H}{1} regions, even if remaining, is well accounted for in the fitted tip-contrast ratio relationship to the degree that the fields are positioned similarly to those used here. We also note that Type Ia supernovae hosts do not have \ion{H}{1} maps available or have \ion{H}{1} maps that lack sufficient resolution \citep[SN hosts $\sim$20 Mpc, see][]{Condon_1987ApJS...65..485C, Condon_1996ApJS..103...81C}. Spatial clipping is a method that can remove contamination from younger stars that can be applied to both anchors and host galaxies as opposed to direct removal of \ion{H}{1} regions via \ion{H}{1} maps.

The tip-contrast ratio relationship represents an approach that is distinct but consistent with previous measurements of the tip and can be used to explain the large ($\sim 0.1$~mag) spread in recent tip zero-point calibrations \citep[see][]{Blakeslee_2021ApJ...911...65B, Freedman_Tensions_2021ApJ...919...16F, Li_2022arXiv220211110L}. Instead of assuming that a single, universal tip zero-point that can be used with any stellar population regardless of its composition of different stellar and luminosity function properties or measurement setup, we standardize the tip and its zero-point to account for the variations due to these astrophysical properties and measurement parameters such as age and smoothing. This interpretation also means that multiple tip discontinuities can exist in a single luminosity function, as fields with mixtures of stars with different stellar properties can have multiple sub-populations that exhibit a slightly different tip, though detecting these peaks will depend on the relative distance between the peaks and the degree of smoothing and noise. Previous measurements of the tip often selected a single tip discontinuity over others or proximity to an accepted range of tips based on previous studies and required a supervision to determine which peak to choose that rejects certain tips in favor or others. We discuss an example of this in Appendix \ref{sec:LMC}. The CATs algorithm used in this study is unsupervised to avoid this potential subjective bias. .

We hypothesize the TCR relationship originates from a combination of astrophysical and measurement processes such as age, metallicity, photometric errors, and smoothing.  Observationally, we find that the tip varies with the contrast ratio similar to the trends found in \cite{Anderson_2023arXiv230304790A} for the LMC and \cite{Wu_2022arXiv221106354W} for GHOSTS galaxies and argue that a given tip zero-point should be used to calibrate tip measurements of similar characteristics. Even if the TCR were entirely due to a particular choice of measurement setup and has no astrophysical origin, or vice versa, it does not change the fact that there is an observable variation in the tip against a parameter, the contrast ratio, that was previously not accounted for and that this variation should be accounted for in future measurements involving the tip to ensure greater consistency when using the tip to measure distances. There may be an additional uncertainty that will need to be propagated into distance measurements due to a change in the slope of the TCR due to different photometric noise in different galaxies, for instance. We plan to investigate this topic in further detail and note that this does not change the results presented in this paper, which calibrates the TCR in a single galaxy.

The tip currently plays an important role in the Hubble Tension debate as it can be used to measure the $H_0$ independently and parallel to the Cepheid distance scale. It will be essential to measure the TRGB-based $H_0$ using this improved standardization to increase the accuracy of future extragalactic distance measurements. We present this work in a companion paper in \cite{Scolnic_2023arXiv230406693S}.

\section{Data Availability}
\label{sec:Data_Availability}
The data and CATs algorithm used for this analysis can be found at \url{https://github.com/JiaxiWu1018/CATS-H0}. We encourage the community to reproduce and provide feedback on this analysis.

\appendix

\section{Field 4 Groups}
\label{sec:Field4_Groups}
\begin{figure}[H]
\epsscale{1}
\plotone{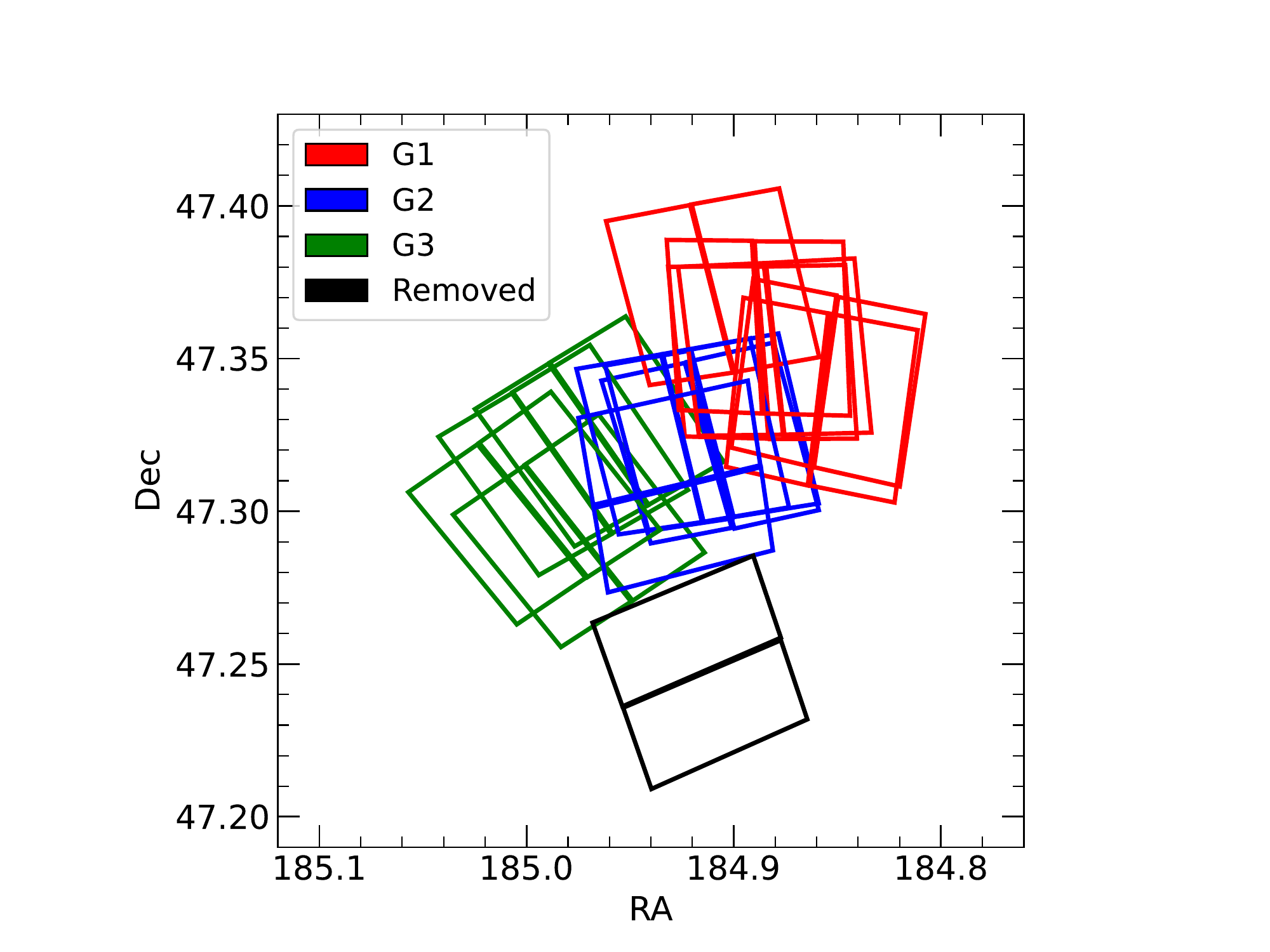}
\caption {Grouping of the three G-fields for Field 4. We remove the fields colored in black and green; Field 4 comprises of only the fields colored in blue and red. In our analysis we measure the tip separately for G1 and G2.}
\label{fig:G_Groups}
\end{figure}

We show the positions of the three groups that make up Field 4 in Fig. \ref{fig:G_Groups}. 

\section{Measuring Tips}
\label{sec:Measuring_Tips}
 
The CATs algorithm \citep{Wu_2022arXiv221106354W} measures the spatial density of blue stars and masks high density regions to remove young-star contamination (a step called spatial clipping).   It then selects a diagonal region in the CMD of fixed width and variable slope and position to maximize the number of RGB stars contained within that region.  Finally, it bins and smooths the luminosity function consisting of red giants and asymptotic giants.  We plot the color magnitude diagrams for the 10 fields analyzed in this study after the photometric quality cuts and spatial cuts in Fig. \ref{fig:CMDs}, which also shows the diagonal regions selected to measure the LFs.  

\begin{figure}{}
\epsscale{1}
\plotone{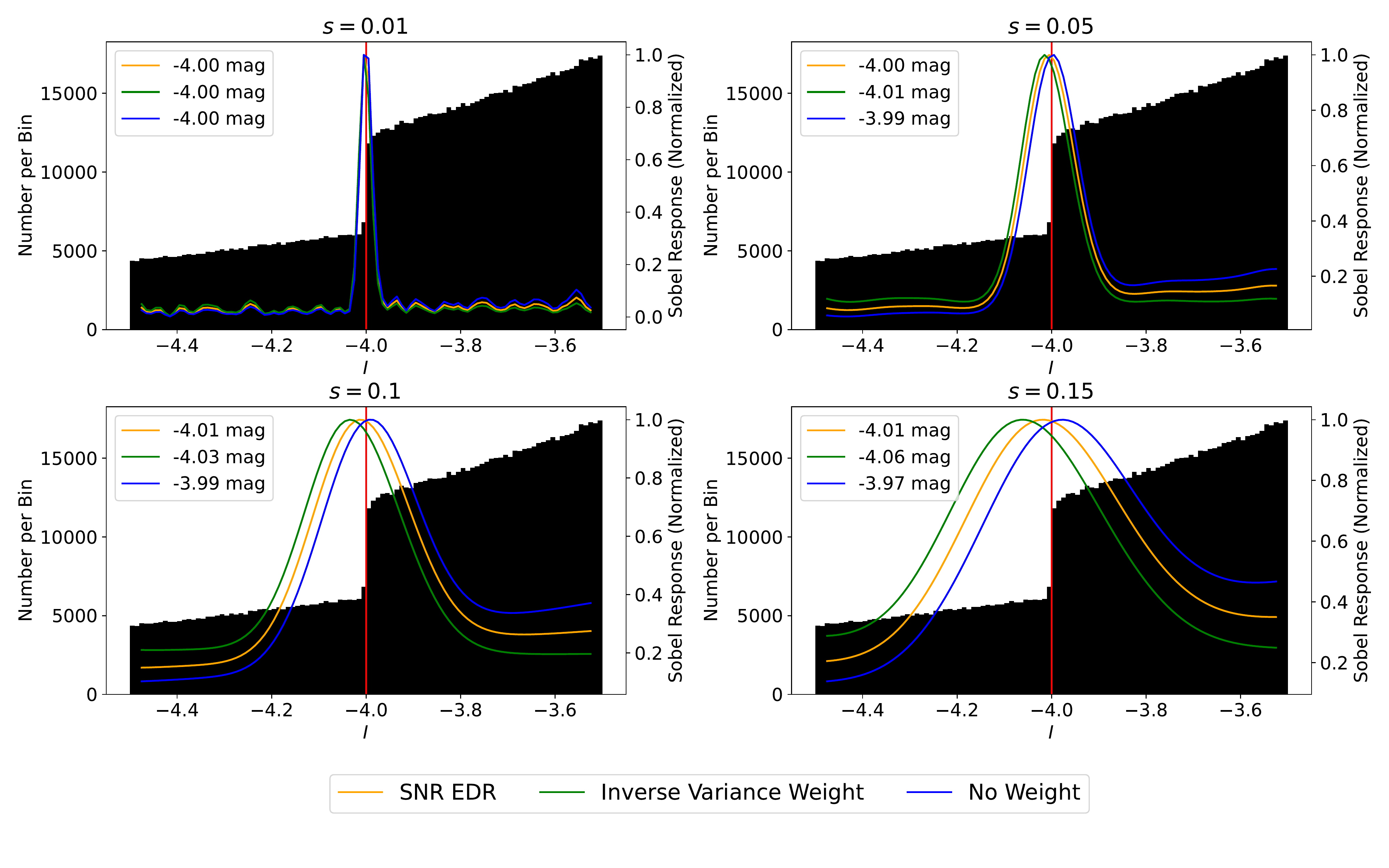}
\caption {Sobel responses with four smoothings (s = 0.01 (top left), 0.05 (top right), 0.1 (bottom left), and 0.15 (bottom right)) and three different weightings (CATs (orange; Eq. \ref{eq:Error_Weighting}), inverse variance (green; Eq. \ref{eq:Inverse_Variance}), and none (blue)) on the same luminosity function simulated with 1,000,000 samples using von Neumann rejection sampling and a TRGB at $-4$~mag and shape parameters of  $\alpha = \beta 
= \gamma = 0.3$ for the broken-power law model described in \cite{Mendez_2002AJ....124..213M,Makarov_2006AJ....132.2729M, Li_2022arXiv220211110L}. The vertical red lines mark the location of the TRGB in each subplot.}
\label{fig:Various_Smoothings_Weights}
\end{figure}

The LFs appear rather fuzzy near the visual tip of the TRGB.  Smoothing and weighting are necessary to evaluate the derivative of such noisy data. However, we note that various options used to smooth and weight the derivative will generate a bias in the determination of its position, even without noise, due to the asymmetric shape of the LF.  This is illustrated in Figure \ref{fig:Various_Smoothings_Weights} where we show several common choices.  

 In Fig. \ref{fig:Various_Smoothings_Weights}, we simulate a luminosity function with 1,000,000 samples using von Neumann rejection sampling and a broken power law model described in \cite{Mendez_2002AJ....124..213M, Makarov_2006AJ....132.2729M, Li_2022arXiv220211110L} using realistic model parameters of $m_{TRGB} = -4$~mag, $\alpha = \beta 
= \gamma = 0.3$. We Gaussian (GLOESS) smooth the LF at values of $s$ = 0.01, 0.05, 0.1, and 0.15 and evaluate the derivative of the LF with three choices of functions with different weight or normalization:

Signal-to-Noise EDR \citep{Wu_2022arXiv221106354W, Gorski_2018AJ....156..278G}:
\begin{equation}
\label{eq:Error_Weighting}
    EDR(i) = \frac{N(i + 1) - N(i - 1)}{\sqrt{N(i + 1) + N(i - 1)}}
\end{equation}

Inverse Variance Weighting:
\begin{equation}
\label{eq:Inverse_Variance}
    EDR(i) = \frac{N(i + 1) - N(i - 1)}{N(i+1) + N(i - 1)} 
\end{equation}

No Weighting:
\begin{equation}
\label{eq:No_Weighting}
    EDR(i) = N(i + 1) - N(i - 1)
\end{equation}

 We note that the signal-to-noise EDR in Equation \ref{eq:Error_Weighting} is different than that used in \cite{Hatt_2017ApJ...845..146H}, which applies the SNR as a weight on top of the Sobel filter (see their Section 3.3.2). We plot the edge detection responses from these variations in Fig. \ref{fig:Various_Smoothings_Weights}. We find that adopting different smoothing and weighting can shift the measured TRGB by several hundredths of a magnitude.

To reduce this bias along the distance ladder, we have chosen the EDR in Eq. \ref{eq:Error_Weighting} which has the smallest bias, and we set the smoothing value to be the same in both this calibration, NGC 4258, and application in SN hosts, for first-order cancellation.  Following the optimal values found in \cite{Wu_2022arXiv221106354W}, we use baseline parameters: spatial clipping of 10$\%$, color band of 1~mag, smoothing of 0.1, Sobel threshold fraction of 0.6, `error weighting' EDR.

\section{Simulated LF with Different Smoothing}
\label{sec:Diff_Smoothing}

\begin{figure*}[ht!]
  \centering
  \begin{tabular}{ccc}
    \includegraphics[width=0.5\linewidth]{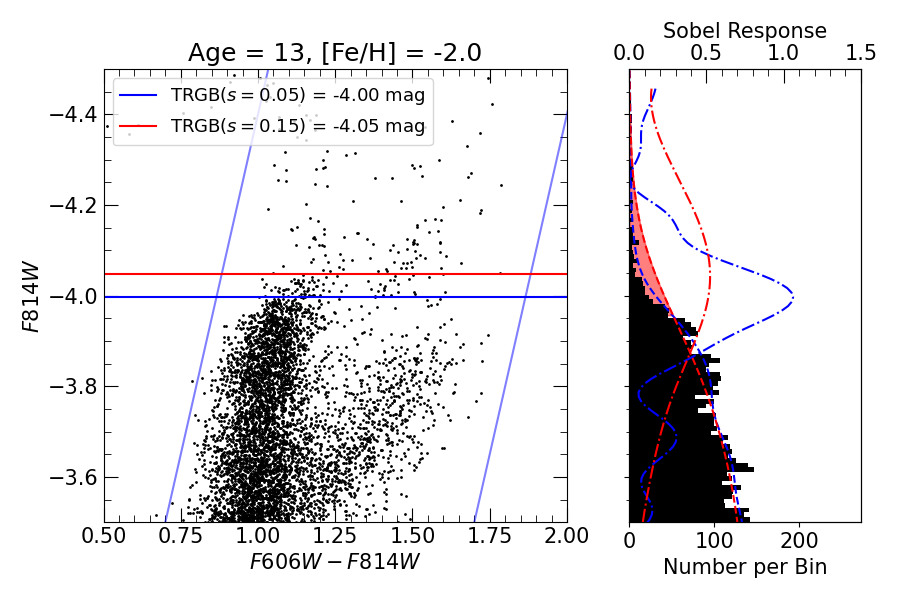} &
    \includegraphics[width=0.5\linewidth]{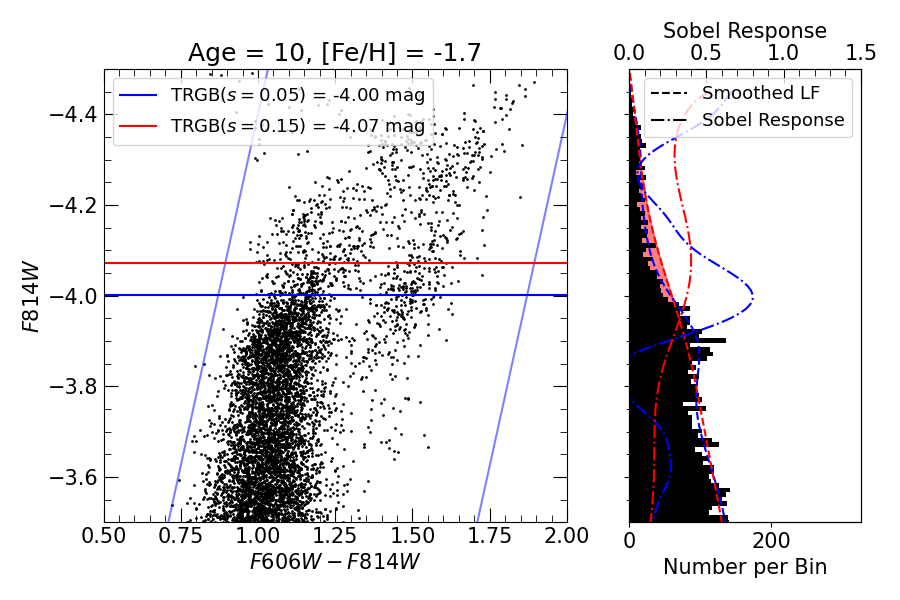} \\
    \includegraphics[width=0.5\linewidth]{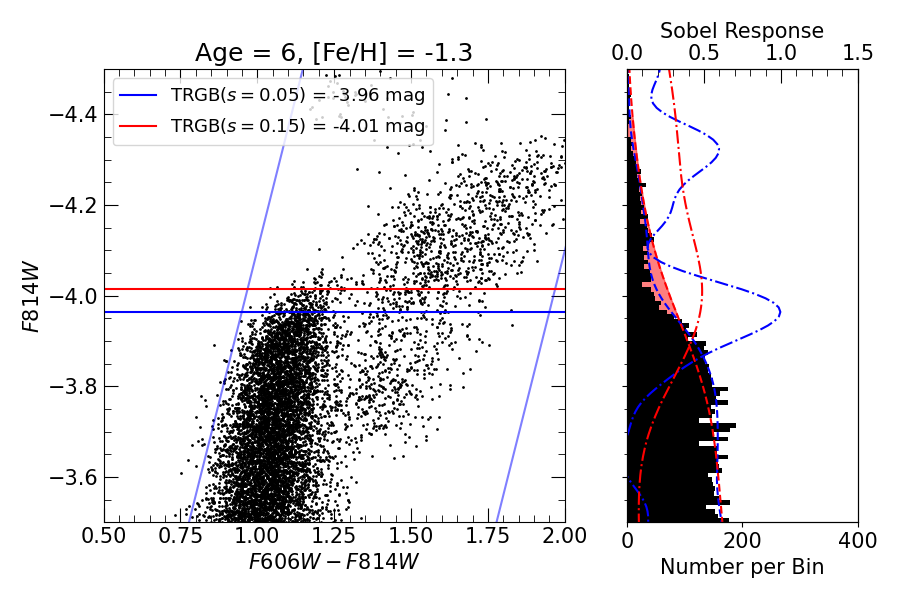} &
    \includegraphics[width=0.5\linewidth]{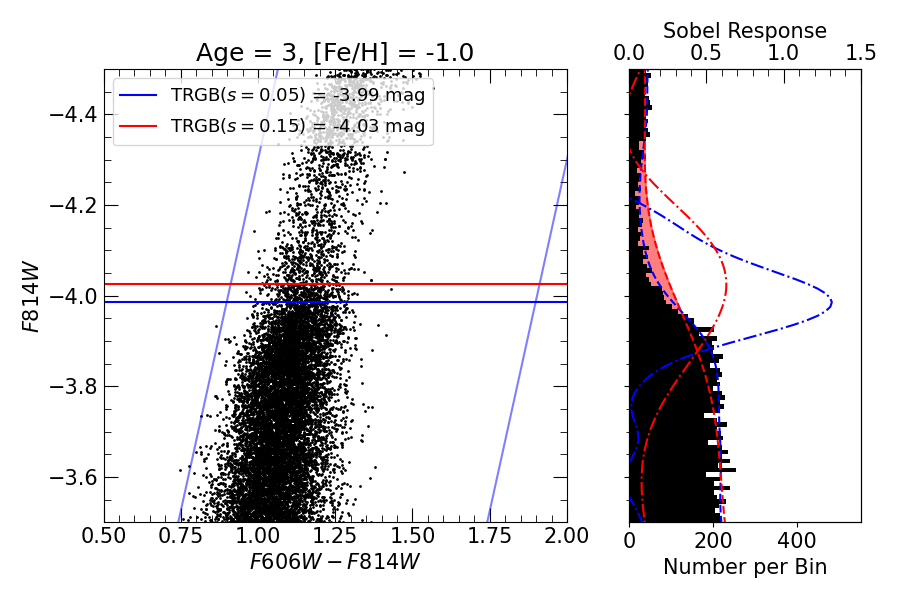} \\
  \end{tabular}
\caption{CMDs and luminosity functions for simulated stellar populations of [Fe/H] = $−2.0$, $−1.7$, $−1.3$, and $−1.0$ and ages of 13, 10, 6, and 3 Gyr, respectively, from clockwise order from the top left. Blue diagonal lines in the left subpanels are the color bands used. Red and blue lines in both subpanels correspond to smoothing values of $s = 0.05$ and $s = 0.15$, respectively. Solid, dashed, and dot-dashed lines correspond to the measured tip, smoothed luminosity function (LF), and Sobel response, respectively. We shade the extra space representing extra stars underneath the smoothed luminosity function created by applying a higher smoothing parameter in red.}
\label{fig:Sim_Sobel_Smoothing}
\end{figure*}

We visualize the effect of smoothing on the measured tip and TCR with Fig. \ref{fig:Sim_Sobel_Smoothing}, where we generate four simulated populations with [Fe/H] = $-2.0, -1.67, -1.33$, and $-1.0$ and ages of 13, 9.7, 6.3, and 3 Gyr, respectively. In the CMDs, the blue slanted lines represent the color band optimized with the CATs algorithm, and the histogram in the right panel correspond to stars within the color band only. The blue and red lines correspond to measurements using smoothing parameters of $s = 0.05$ and 0.15, respectively. Dashed lines correspond to the smoothed luminosity functions, and dot-dashed lines correspond to the Sobel responses. We add the same 0.05~mag noise to the $F814W$ and $F606W$ magnitudes as described above and measure the `measured' tip using the CATs algorithm. From this plot, we see that increasing smoothing from 0.05 to 0.15 pushes stars fainter than the tip across the tip in the brighter direction by flattening the tip discontinuity. This in turn pushes the measured tip in the brighter direction, as shown by the separation between the red horizontal lines in each subplot.

From Fig. \ref{fig:Sim_Sobel_Smoothing}, we observe that 1) adding noise tends to scatter the tip brightwards, as there are more RGB stars that are moved to the region brighter than the tip than vice versa due to the intrinsic relative number of stars in each evolutionary stage, and 2) increasing the smoothing from 0.05 to 0.15 biases the measured tip in the brightwards direction.  The extra stars that are pushed from the region fainter than the measured tip to the region brighter than the measured tip can be visualized as the shaded red space under the dashed lines in each panel. A larger smoothing generates a larger region indicating more stars around the tip are pushed in the brightwards direction. This effect is intensified with larger contrast ratios where the tip discontinuity is larger.  Smoothing values used by CATs typically do not reach this high, however, we caution that oversmoothing can bias the measured tip in the brightwards direction depending on the obseved luminosity function \citep{Anderson_2023arXiv230304790A}. This also illustrates the need for applying consistent smoothing for both zero-point and apparent magnitude tip measurements, as different smoothing values will offset the measured tip from the true tip by different amounts due to this effect. A similar effect was found by \cite{Anderson_2023arXiv230304790A} who found a bias that is not necesarily linear and is a function of contrast (with a different definition than that used here) and steepness of the tip discontinuity for a fixed smoothing value. They note that there is often a lack of objective criterion from choosing the smoothing scale (for instance, \cite{Hatt_2017ApJ...845..146H} suggests that one can increase the smoothing scale until there is a single, dominant, local maximum in the Sobel response) and define a criterion such that the choice of smoothing should not change the result more than 0.1 mag/mag. They also note that the smoothing scale should be used consistently in the anchor and Type Ia supernovae calibrator galaxies as this bias would otherwise not cancel out. 

As the tip is biased in the brighter direction for a given luminosity function, the decrease in the number of AGB stars is greater than the decrease in the number of RGB stars in the 0.5~mag region around the tip. In terms of a broken power law model \citep{ Mendez_2002AJ....124..213M, Makarov_2006AJ....132.2729M, Li_2022arXiv220211110L}, the $c$ or $\gamma$ parameter (AGB slope) is greater than the $a$ or $\alpha$ parameters (RGB slope). As the tip is biased brightwards with larger smoothing, R will also change accordingly, given the same luminosity function is used in this comparison. We note this situation, where the luminosity function remains constant but the measured tip changes, is different from the tip-contrast ratio calibration where the population of stars changes for each measurement.

\section{LMC}
\label{sec:LMC}

In this section, we investigate the feasibility of using the tip calibration from \cite{Hoyt_2023NatAs.tmp...58H} for $H_0$ measurements and whether the TCR is consistent with their findings. \cite{Hoyt_2023NatAs.tmp...58H} measure a sub-percent calibration of the tip using red giants in the LMC by selectively removing 15 out of 20 fields on the basis of a subjective evaluation of their Sobel responses and bootstrap asymmetry. We note that the fields \cite{Hoyt_2023NatAs.tmp...58H} removes are systematically fainter, suggesting a relationship between the tip magnitude and the cleanness of the Sobel response.

\begin{figure}[H]
\epsscale{1}
\plotone{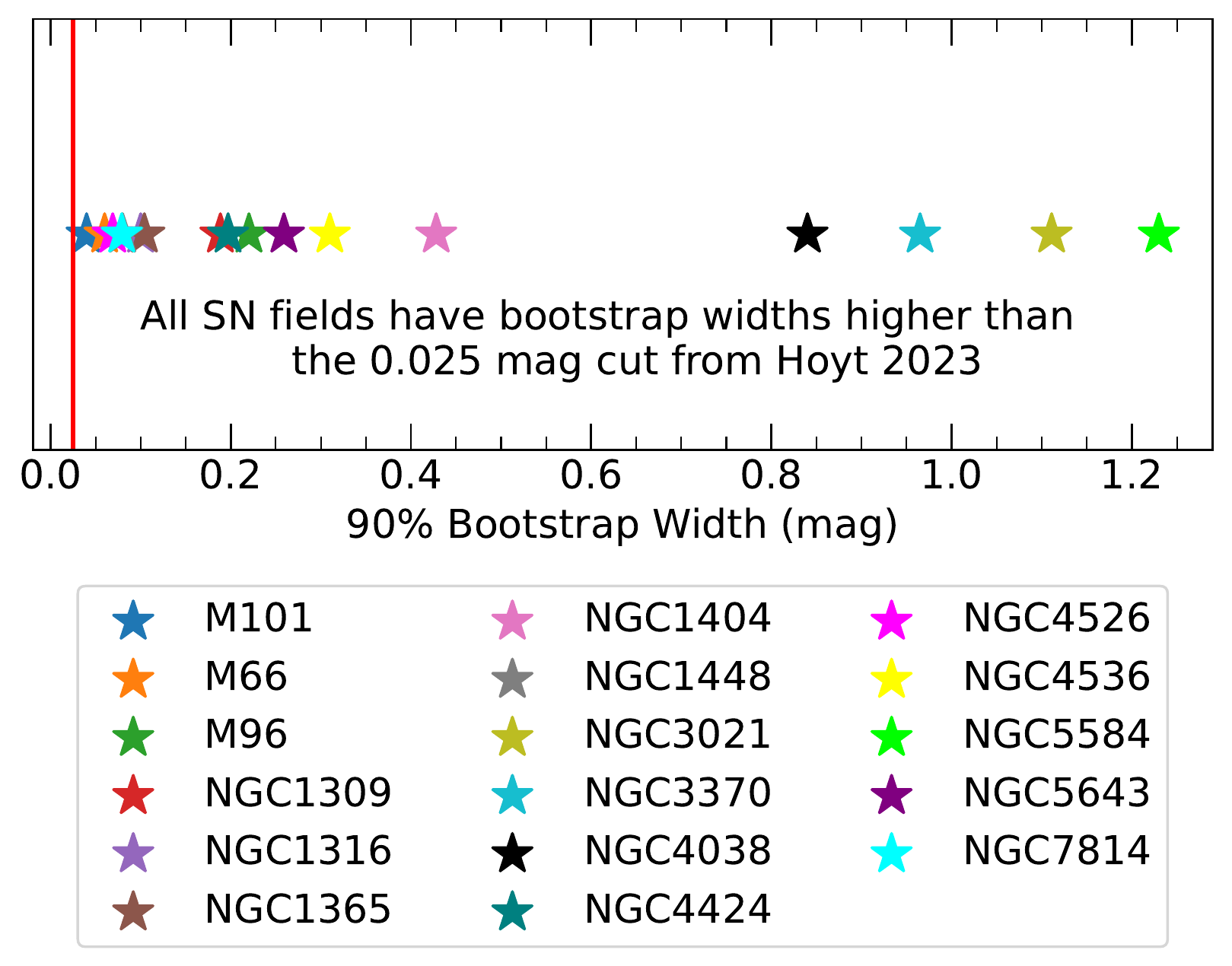}
\caption {90$\%$ bootstrap widths for the SNIa host galaxies from \cite{Scolnic_2023arXiv230406693S}. We indicate the limit of 0.025~mag used by \cite{Hoyt_2023NatAs.tmp...58H} to select LMC fields such that all bootstrap widths greater than 0.025~mag would be rejected. We find that none of the SNIa host galaxies would pass the cuts made by \cite{Hoyt_2023NatAs.tmp...58H} for the LMC.}
\label{fig:LMC_Bootstrap}
\end{figure}

To see if these same cuts can be feasibly applied to Type Ia SN hosts, we first apply the CATs algorithm to the Type Ia host galaxies from \cite{Scolnic_2023arXiv230406693S}. We perform 10,000 bootstrap resamples for each galaxy and plot the 90$\%$ width of the boostrapped samples in Fig. \ref{fig:LMC_Bootstrap}.  We record only the tip corresponding to the highest Sobel peak. 

We find that none of the galaxies would pass the cuts show in Extended Data Fig. 1 shown in \cite{Hoyt_2023NatAs.tmp...58H}. In other words, it would not be possible to selectively choose brighter fields in these host galaxies as done in \cite{Hoyt_2023NatAs.tmp...58H} for the LMC. We conclude that the tip calibration from \cite{Hoyt_2023NatAs.tmp...58H} cannot be used to calibrate tip measurements in host galaxies to measure $H_0$.

\begin{figure}[H]
\epsscale{1}
\plotone{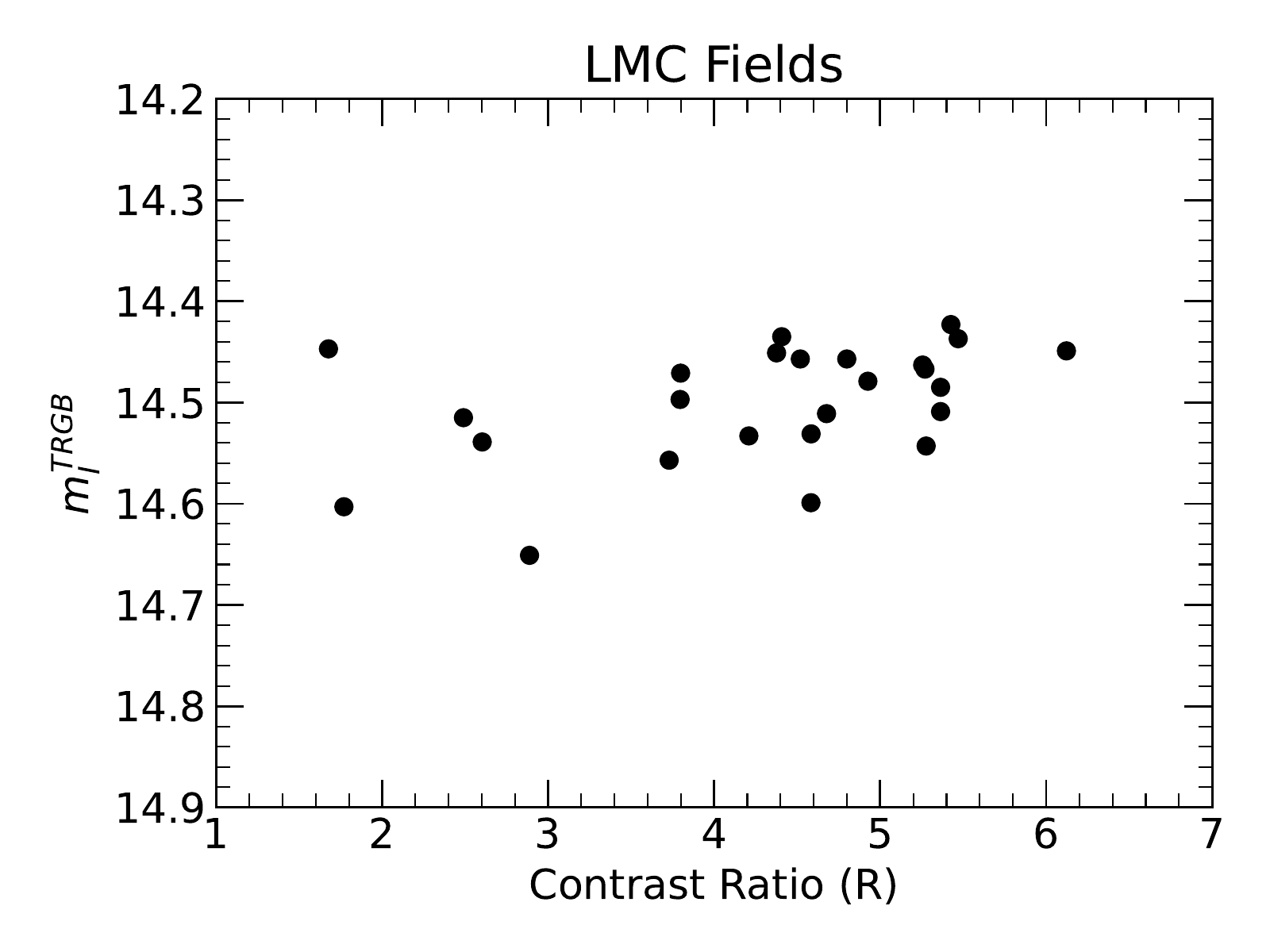}
\caption {Tips from \cite{Hoyt_2023NatAs.tmp...58H} as a function of their contrast ratios.}
\label{fig:LMC_TCR}
\end{figure}

Next, we take the same 20 measured tips from \cite{Hoyt_2023NatAs.tmp...58H}, which were measured in 20 Voronoi fields based on the concepts from \citep{Cappellari_2003MNRAS.342..345C,Hoyt_2018ApJ...858...12H} and attributed to M. Seibert, and measure their contrast ratios. We plot the tips as a function of their contrast ratios in Fig. \ref{fig:LMC_TCR}. We see a trend that is indicative of a TCR. We do not plot uncertainties or apply a fit because the errors quoted in \cite{Hoyt_2023NatAs.tmp...58H} do not account for a field-to-field dispersion such as the one calibrated in \cite{Wu_2022arXiv221106354W}. In addition, the error equation in Eq. \ref{eqn:error} was calibrated using fainter GHOSTS galaxies and will overestimate the uncertainties for the LMC tips, which are much brighter. We defer a more detailed treatment with the CATs algorithm to future work.

\begin{acknowledgments}
SL is supported by the National Science Foundation Graduate Research Fellowship Program. DS is supported by Department of Energy grant DE-SC0010007, the David and Lucile Packard Foundation, the Templeton Foundation and Sloan Foundation. RIA is funded by the SNSF through an Eccellenza Professorial Fellowship, grant number PCEFP2\_194638.
RLB acknowledges support from NSF-AST 2108616. 

 This research has made use of NASA’s Astrophysics Data System and is based on observations made with the NASA/ESA Hubble Space Telescope obtained from the Space Telescope Science Institute, which is operated by the Association of Universities for Research in Astronomy, Inc., under NASA contract NAS 5–26555. These observations are associated with programs 9477, 10399, 16198, 16688, and 16743.
\end{acknowledgments}

\bibliography{MAIN_DOCUMENT}{}
\bibliographystyle{aasjournal}

\end{document}